\documentclass[usenatbib]{mn2e}
\usepackage{aas_macros}
\usepackage{graphicx}
\usepackage{amssymb}
\usepackage{amsmath}
\usepackage{enumerate}

\usepackage[usenames,dvipsnames]{color}
\usepackage[normalem]{ulem}

\newcommand{\der}{\ensuremath{{\rm d}}}

\newcommand{\eqn}[1]{equation~\eqref{#1}}

\newcommand{\fig}[1]{Figure~\ref{#1}}

\newcommand{\avg}[1]{\ensuremath{\left\langle \,#1\, \right\rangle}}

\newcommand{\be}{\begin{equation}}
\newcommand{\ee}{\end{equation}}
\newcommand{\bear}{\begin{eqnarray}}
\newcommand{\ear}{\end{eqnarray}}

\newcommand{\f}{\frac}

\definecolor{darkgreen}{rgb}{0,0.35,0}

\newcommand{\lya}{Ly$\alpha$ }

\begin{document}
\title[LAEs gone missing: evidence for late reionization?]
{Lyman-$\alpha$ emitters gone missing: evidence for late reionization?}

\author[Choudhury et al.]
{\parbox{\textwidth}{Tirthankar Roy Choudhury$^1$,  Ewald Puchwein$^2$, Martin G. Haehnelt$^2$ and \\ James S. Bolton$^3$}  \vspace{3mm}
\\$^1$National Centre for Radio Astrophysics, Tata Institute of Fundamental Research, Pune 411007, India
\\$^2$Institute of Astronomy and Kavli Institute for Cosmology, University of Cambridge, Madingley Road, Cambridge CB3 0HA, UK
\\$^3$School of Physics and Astronomy, University of Nottingham, University Park, Nottingham, NG7 2RD, UK} 

\maketitle

\date{\today}

\begin{abstract}

We combine high resolution hydrodynamical simulations with an
intermediate resolution, dark matter only simulation and an analytical
model for the growth of ionized regions to estimate the large scale
distribution and redshift evolution of the visibility of \lya emission
in $6 \leq z\leq 8$ galaxies. The inhomogeneous distribution of
neutral hydrogen during the reionization process results in
significant fluctuations in the \lya transmissivity on large scales. 
The transmissivity  depends  not only on the ionized fraction of the intergalactic
medium by volume and the amplitude of the local ionizing background, but
is also rather sensitive to  the evolution of the relative velocity shift of the \lya emission line due to
resonant scattering.
We reproduce a decline in the space density of \lya emitting
galaxies as rapid 
as observed with a rather rapidly evolving neutral fraction between $z = 6$ -- $8$,
and a typical \lya line velocity offset of $100\rm\,km\,s^{-1}$ redward of systemic at $z=6$ which
decreases toward higher redshift.
The new (02/2015) Planck results
indicate such a recent end to reionization is no longer disfavoured by
constraints from the cosmic microwave background.

\end{abstract}

\section{Introduction}
\label{outline}

Colour selection techniques applied to deep imaging surveys with
Advanced Camera for Surveys on the Hubble Space Telescope have enabled
the routine identification of large numbers of high redshift galaxies
out to redshifts as high as $z=10$, deep in the epoch of hydrogen
reionization \citep[for recent results,
  see][]{2013ApJ...777L..19L,2013ApJ...773...75O,2013ApJ...765L..16B,2015ApJ...803...34B,2014ApJ...786..108O,2014arXiv1410.5439F,2014arXiv1412.1472M}. The
evolution of the galaxy luminosity function inferred from these
surveys thereby appears smooth with a rather modest decrease of the
space density of continuum emission selected galaxies with increasing
redshift.

This rather smooth evolution of continuum selected galaxies is,
however, not mirrored by the \lya emission from high redshift
galaxies, which appears to decline rather rapidly at $z>6$. This was
first noted in \lya emission selected galaxy samples
\citep{2006ApJ...648....7K} and has been since confirmed by several
\lya emitter surveys
\citep{2010ApJ...723..869O,2010ApJ...725..394H,2011ApJ...734..119K,2014ApJ...797...16K}. It
has also been established for continuum selected galaxies by a careful
investigation of their \lya equivalent width distribution
\citep{2010ApJ...725L.205F,2011ApJ...728L...2S,2011ApJ...743..132P,2012ApJ...744..179S,2012ApJ...747...27T,2012MNRAS.427.3055C,2014ApJ...793..113P,2014MNRAS.443.2831C,2014ApJ...795...20S,2014ApJ...794....5T}.
This rapid evolution of the \lya emission in high redshift galaxies
was initially surprising; reionization models had predicted that the
intergalactic medium (IGM) needed to still be substantially neutral
(with volume averaged neutral fractions $>50$ per cent at $z \sim 7$)
for not yet reionized gas to strongly affect the visibility of \lya
emission from high-redshift galaxies
\citep{2011ApJ...743..132P,2012ApJ...744...83O,2012ApJ...744..179S,2013MNRAS.428.1366J,2014ApJ...797...16K,2014ApJ...788...87F}.

There is, however, now a general consensus that the Universe is mostly
ionized at $z\sim 6$, based on detailed analyses of QSO absorption
spectra
\citep{2006AJ....132..117F,2006MNRAS.370.1401G,2008MNRAS.386..359G,2012MNRAS.421.1969R,2015MNRAS.447..499M}.
Both CMB data and the ionizing emissivity inferred from QSO absorption
spectra and high-redshift galaxy surveys also suggest that large
neutral hydrogen fractions of fifty per cent or more are only expected
to be present at redshift $z>7$
\citep{2010MNRAS.408...57P,2011MNRAS.413.1569M,2012MNRAS.423..862K,2012MNRAS.419.1480M,2013ApJ...768...71R}. It
has therefore been argued that the rapid evolution of the \lya
emission in high-redshift galaxies may be due to a rapid evolution of
the internal properties of high redshift galaxies with regard to the
escape of \lya emission \citep{2012MNRAS.421.2568D}. While this could
certainly explain the rapid demise of \lya emission, there are,
however, concerns about the plausibility of such a rapid evolution
within the smoothly evolving hierarchical paradigm for galaxy
formation suggested by the $\Lambda$CDM cosmological
framework. Alternatively, \citet{2014MNRAS.440.3309D} have argued that
an evolving Lyman continuum escape fraction may in part explain the
observed \lya emitter (LAE) evolution, while
\citet{2014MNRAS.437.2542T} note that sample variance could also play
a role due to the patchy nature of reionization.

Motivated by these findings, along with the possible discovery of a
damping wing that extends redwards of the systemic host galaxy
redshift in an (absorption) spectrum of the $z=7.085$ QSO ULAS
J1120+0641 \citep{2011Natur.474..616M,2011MNRAS.416L..70B},
\citet{2013MNRAS.429.1695B} performed detailed numerical modelling of
the \lya opacity during the epoch of reionization.  Their simulations
employed sufficient resolution to account for the opacity due to
self-shielded residual neutral hydrogen in the already mostly ionized
regions of the Universe.  In this way, \citet{2013MNRAS.429.1695B}
demonstrated that self-shielded neutral regions in the late stages of
reionization can lead to substantial \lya opacity even for volume
filling factors of neutral hydrogen as low as ten percent, potentially
explaining the observed rapid decline of LAEs.  However, the
simulations of \citet{2013MNRAS.429.1695B} were performed with a
volume too small to investigate the spatial distribution and evolution
of the intervening \lya opacity. They were furthermore based on a
simple backwards extrapolation of photoionization rates measured at
lower redshift, and lacked a realistic model for the growth of ionized
regions.  The most recent 
Planck results \citep{2015arxiv150201589P} also now appear to
be fully consistent with models predicting reionization to complete
not much before $z\sim6$, somewhat relaxing the need for low ($\sim
10$ per cent) neutral hydrogen fractions at $z \sim 7$.

To investigate this further we therefore combine high-resolution
hydrodynamical simulations which reproduce the observed IGM \lya
opacity at $2<z<6$ with an analytical model for the growth of ionized
regions built on a large, dark matter only simulation.  Our analytical
model for the growth of ionized regions is based on that described in
\citet{2007ApJ...654...12Z}, \citet{2009MNRAS.394..960C} and
\citet{2012MNRAS.426.3178M}, and has been successfully tested against
cosmological radiative transfer simulations
\citep{2006NewA...11..374M,2012MNRAS.423.2222I} in
\citet{2014MNRAS.443.2843M}. The hybrid approach we employ here is
thus similar to that presented in \citet{2015MNRAS.446..566M}, but
note that we calibrate our model for the growth of ionized regions to
match observations of photoionization rate
\citep{2011MNRAS.412.1926W,2011MNRAS.412.2543C} and mean free path of
ionizing photons \citep{2010ApJ...721.1448S,2014MNRAS.445.1745W} at $5
\lesssim z \lesssim 6$. Our default model for the evolution of the
volume filling factor of ionized regions is thereby chosen to be close
to that predicted by the recent update of the
\citet{2012ApJ...746..125H} metagalactic UV background model (HM2012
hereafter), which incorporates a wide range of recent observational
constraints.  However, we shall also explore the effect of
reionization completing somewhat later than in the HM2012 model.
We use our hybrid technique not only to make predictions for the
probability distribution of \lya transmissivity due to the
increasingly neutral, inhomogeneously reionized intervening IGM, but
we also investigate the large-scale spatial distribution of the
transmissivity.

The paper is structured as follows.  In Section 2 we describe our
numerical simulations, and in Section 3 we discuss the spatial
distribution of neutral hydrogen and the reionization history within
our hybrid simulations. Section 4 outlines our results, including
predictions for the spatial distribution of the \lya emission line
transmissivity under a wide variety of model assumptions. We conclude
in Section 5.  A flat $\Lambda$CDM cosmological model with $\Omega_m =
0.305,~ \Omega_b = 0.048,~ h = 0.679,~ n_s = 0.96,~ \sigma_8 = 0.827$
is adopted throughout \citep{2014A&A...571A..16P}.

\section{The spatial distribution of neutral hydrogen}

\subsection{The need for a hybrid technique}

Simulating the spatial distribution of neutral hydrogen during and
after reionization is a computationally very demanding task.  The
typical sizes of ionized regions before overlap are several tens of
(comoving) Mpc, while the scale of the optically thick regions acting
as sinks of ionizing photons in already ionized regions are only tens
of kpc.  Simulations that follow the growth of ionized regions
therefore require box sizes of hundred cMpc or more, while modelling
the \lya opacity correctly requires high-resolution hydrodynamical
simulations for which achievable box sizes are typically ten $h^{-1}$ cMpc.

In order to achieve this dynamic range, we therefore employ a hybrid
technique where we calculate the growth of ionized regions with a well
tested analytical method which relates the growth of structure as
probed by collapsed dark matter haloes to the production of ionizing
photons
\citep{2007ApJ...654...12Z,2009MNRAS.394..960C,2012MNRAS.426.3178M,2014MNRAS.443.2843M}.
We apply this technique to a $1200^3$ particle, large ($100 \, h^{-1}$
cMpc) collisionless dark matter simulation with moderate
resolution. The hydrogen distribution in the ionized regions is then
modelled by replicating a smaller $2\times 512^3$ particle, $10 \,
h^{-1}$ cMpc high-resolution hydrodynamical simulation on top of the
ionization structure from the larger box.  In the following, we will
refer to these two models simply as the ``large box'' and ``small
box'' simulations, respectively.  

A $2\times 5120^3$ particle hydrodynamical simulation would have been
required to achieve the same mass resolution within a $100 \, h^{-1}$
cMpc box, which is still beyond even the most ambitious current state-of-the-art hydrodynamical simulations of the IGM
(\citealt{2014MNRAS.444.1518V,2015MNRAS.446.3697L,2015MNRAS.446..521S}).
Note, however, that the hybrid simulation neither captures the large
scale power in the density field nor the correlation between 
the morphology of the large scale ionized regions and the
local density field.  On average, the density of gas and the abundance
of collapsed structures will thus be underestimated in ionized
bubbles; they correspond to dense regions in the large box, but are
populated with small boxes of mean density in the hybrid simulation. Our
models may therefore underestimate the number of absorbers in ionized
bubbles. On the other hand, our hybrid simulation assumes a constant background photoionization rate within the ionized regions, which may lead to overestimation of the number of absorbers close to sources where the actual photoionization rate is expected to be higher than average.

\subsection{Modelling the growth of ionized regions}
\label{sec:largebox}

\begin{figure*}
\centering
\includegraphics[angle=0,width=0.95\textwidth]{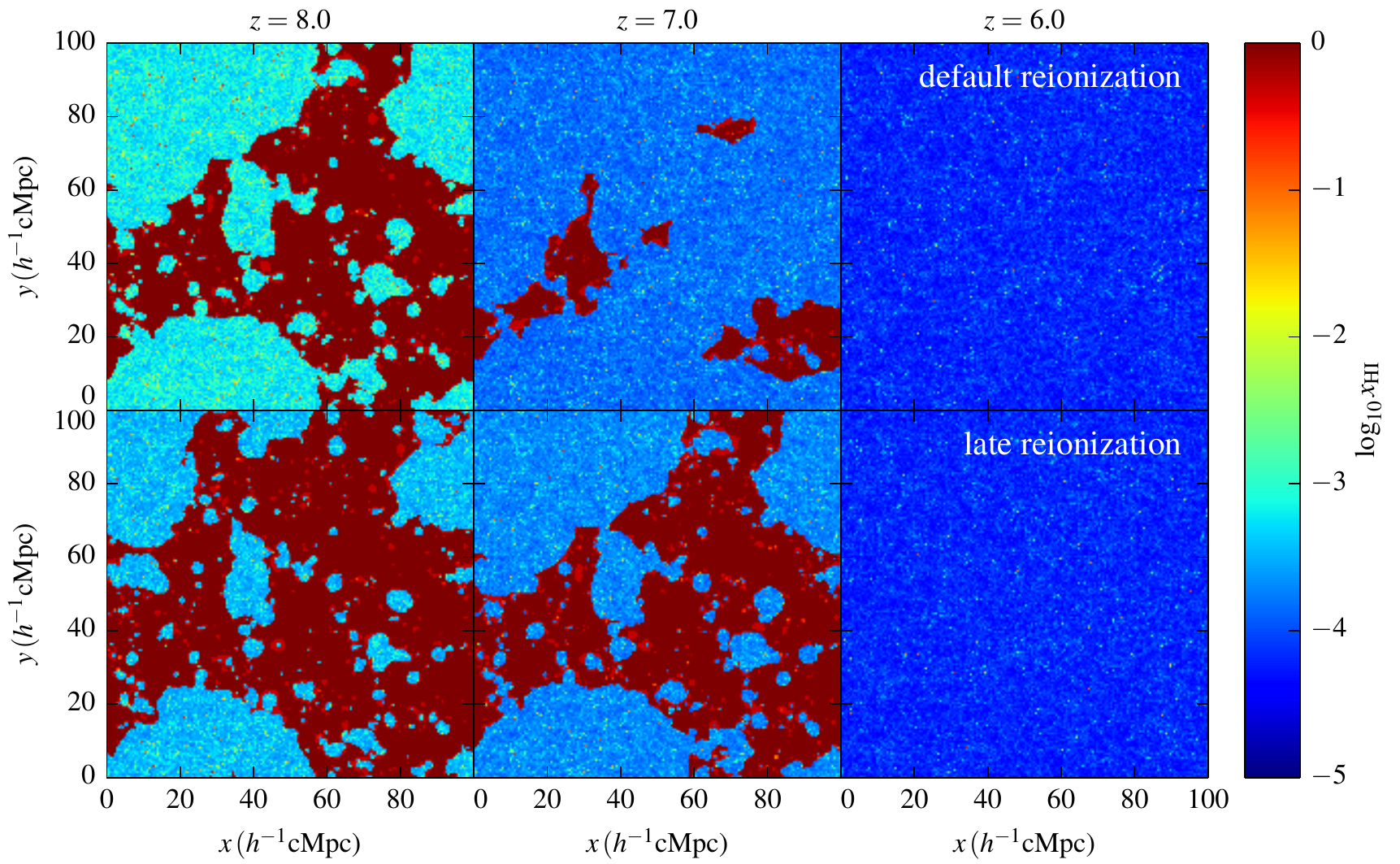}
\caption{Maps of the neutral hydrogen fraction at $z=8, 7$ and $6$,
  obtained with our analytic method for modelling the growth of
  ionized regions within our large simulation box. The \textit{top}
  panels show our default reionization model, while the
  \textit{bottom} panels show our late reionization model.}
\label{fig:plot_ionization_maps}
\end{figure*}

To model the growth of ionized regions we need a model for the
production of ionizing photons. Simple models that relate the
integrated number of ionizing photons in a given region to either dark
matter haloes or the collapsed mass fraction smoothed on a suitable
scale have been shown to reproduce the results of much more demanding,
full radiative transfer simulations
\citep{2006NewA...11..374M,2012MNRAS.423.2222I} remarkably well
\citep{2014MNRAS.443.2843M}. We employ here the model based on
\citet{2007ApJ...654...12Z} and \citet{2009MNRAS.394..960C} and apply it to our
large box simulation, which is dark-matter only, run using the {\sc
  p-gadget3} $N$-body code \citep[the previous version of the code is
  publicly available\footnote{\tt
    http://www.mpa-garching.mpg.de/gadget/} and is described
  in][]{2005MNRAS.364.1105S}.

We first briefly describe the basic steps of this model, which is
similar to that in \citet{2007ApJ...654...12Z} and identical to the
method described in \citet{2009MNRAS.394..960C} (the $\epsilon=0$ case
in their paper) and \citet{2014MNRAS.443.2843M} (the ``sem-num''
model).  As a first step we use a FoF halo finder to identify the
locations and masses of the dark matter haloes in the large box. We
require a group to have at least 32 particles to be labelled as a
halo, thus giving a minimum halo mass of $1.57 \times 10^9 h^{-1}$
M$_{\odot}$. This minimum mass is larger than the typical mass of
atomically cooled haloes with $T_{\rm vir}\sim 10^{4}\rm\,K$, but is
similar to the threshold mass in regions which are affected by
radiative feedback from reionization. Since we are interested in the
late stages of the reionization history where most of the IGM is
heated, the lack of smaller mass haloes in the simulation box should
not impact significantly on our results.

The haloes are assigned emissivities proportional to the halo mass and
are subsequently used for generating the ionization field. A given
location $\mathbf{x}$ in the simulation box is assumed to be ionized
if, within a spherical region of radius $R$ around it, the condition
\be \zeta_{\rm eff} f_{\rm coll}(\mathbf{x}, R) \geq 1,
\label{eq:zeta_eff_barrier}
\ee is satisfied for any value of $R$, where $f_{\rm coll}(\mathbf{x},
R)$ is the collapsed mass fraction within the spherical volume\footnote{Points which do not satisfy condition (\ref{eq:zeta_eff_barrier}) are assigned an ionized fraction $\zeta_{\rm eff} f_{\rm coll}(\mathbf{x}, R_{\rm min})$, where $R_{\rm min}$ is the spatial resolution of the grid which is used for generating the ionization field.}. The
parameter $\zeta_{\rm eff}$ is the effective ionizing efficiency,
which corresponds to the number of photons in the IGM per hydrogen in
stars, compensated for the number of hydrogen recombinations in the
IGM. The large-scale ionization field at a given redshift is thus
determined by only one parameter, $\zeta_{\rm eff}$. Note that we do
not yet model the regions optically thick to ionizing radiation
(``Lyman-limit systems'') while generating the ionization maps from
the large box. We will take care of these separately using the small
box to model the small scale neutral hydrogen distribution assuming an
ionization rate consistent with that used in the large box.

In \fig{fig:plot_ionization_maps} we show maps of the neutral hydrogen
fraction at $z=8$, 7 and 6 at the tail-end of hydrogen reionization
for two models of the evolution of the ionized mass fraction that we
have calibrated to match current observational constraints on the
photoionization rates at $z\leq 6$.  The calibration procedure is
described in detail in Sec.~\ref{sec:calibration}. In the next
section, we proceed to describe our modelling of the optically thick
regions within ionized bubbles using the small box.

\subsection{Modelling the optically thick systems acting as sinks of ionizing photons}
\label{sec:smallbox}

The large box simulation is appropriate for studying large-scale
fluctuations in the ionizing emissivity.  However, it does not resolve
the very small scales which are required to model the optically thick
absorption systems (i.e., the (super) Lyman-limit systems).  Modelling
such systems requires simulations with spatial resolution as small as
$\sim 10$ ckpc. We have thus used a smoothed particle hydrodynamics
simulation using {\sc p-gadget3} with a box size of $10 \, h^{-1}$
cMpc and $2\times 512^3$ dark matter and gas particles, similar to the
model used in \citet{2013MNRAS.429.1695B}. 

We model the distribution of optically thick absorption systems in the
small box as follows. Given the value for the background
photoionization rate in the IGM, $\Gamma_{\rm HI}$, one can obtain the
neutral hydrogen fraction at any location assuming photoionization
equilibrium. However, although this approach is reasonably accurate in
modelling the low density optically thin IGM, it is not appropriate
for self-shielded regions. To account for this in the past it has
often been assumed that the gas remains neutral above a density
threshold $\Delta_{\rm ss}$. Assuming the size of the system to be
given by the Jeans scale \citep{2001ApJ...559..507S}, the threshold
density is then,
\begin{align}
\Delta_{\rm ss} = \,\, &36 \left(\frac{\Gamma_{\rm HI}}{10^{-12} \mathrm{s}^{-1} }\right)^{2/3}~\left(\frac{T}{10^4 \mathrm{K}}\right)^{2/15} \nonumber \\
&\times \left(\f{\mu}{0.61}\right)^{1/3}
\left(\f{f_e}{1.08}\right)^{-2/3}
 \left(\f{1+z}{8}\right)^{-3},
\label{eq:delta_ss}
\end{align}
where $T$ is the gas temperature, $\mu$ is the mean molecular weight
and $f_e = n_e/n_H$ is the ratio of free electrons to hydrogen. The
threshold can be computed at every location in the simulation box
using the densities and temperatures of the gas particles. According
to \eqn{eq:delta_ss}, low values of $\Gamma_{\rm HI}$ at high redshift
can make $\Delta_{\rm ss}$ unrealistically small.  In the models we
use in this paper, this starts to affect our analysis at $z > 8$. To
avoid this, we impose a lower limit of $\Delta_{\rm ss} \geq 2$ on the
self-shielding threshold.

In this work we consider three different models; one that neglects
self-shielding, one with a simple density threshold for self-shielding
(i.e. Eq.~\ref{eq:delta_ss}) and one where self-shielding is
implemented with a fitting formula based on full radiative transfer
simulations.  These are summarized as follows:

\begin{itemize}

\item {\bf No-SS:} self-shielding is neglected in this model.  The IGM
  is assumed to be optically thin and in photoionization equilibrium
  at every location within the ionized regions found in the large box,
  irrespective of the local density.\\

\item {\bf SS:} Hydrogen is assumed  to be completely neutral if $\Delta_H > \Delta_{\rm ss}$, i.e.,
\be
\Gamma_{\rm HI}^{\rm local} = \left\{
\begin{array}{cc}
\Gamma_{\rm HI} & {\rm when}~\Delta_H \leq \Delta_{\rm ss},\\
0 & {\rm if}~\Delta_H > \Delta_{\rm ss},
\end{array}
\right.
\label{eq:ss-b}
\ee
where $\Delta_H$ is the hydrogen overdensity.\\

\item {\bf SS-R:} The SS model predicts a transition in the ionization
  state around the threshold density which is sharper than found in
  radiative transfer simulations involving recombinations
  \citep{2013MNRAS.430.2427R}. In this third model we therefore use
  the empirical fit provided by \citet{2013MNRAS.430.2427R} to their
  radiative transfer calculations, where:\footnote{The relation we
    have used for estimating the self-shielding threshold,
    Eq.~(\ref{eq:delta_ss}), differs slightly from that used in
    \citet{2013MNRAS.430.2427R}, particularly the $T$-dependence of
    $\Delta_{\rm ss}$. However, the difference in the predicted
    photoionization rate is not more than $\sim 10$ per cent.} \be
  \f{\Gamma_{\rm HI}^{\rm local}}{\Gamma_{\rm HI}} = 0.98 \left[1 +
    \left(\f{\Delta_H}{\Delta_{\rm ss}}\right)^{1.64}\right]^{-2.28} +
  0.02 \left[1 + \f{\Delta_H}{\Delta_{\rm ss}}\right]^{-0.84}.
\label{eq:ss-r}
\ee

\end{itemize}

\begin{figure*}
\centering
\includegraphics[angle=0,width=0.95\textwidth]{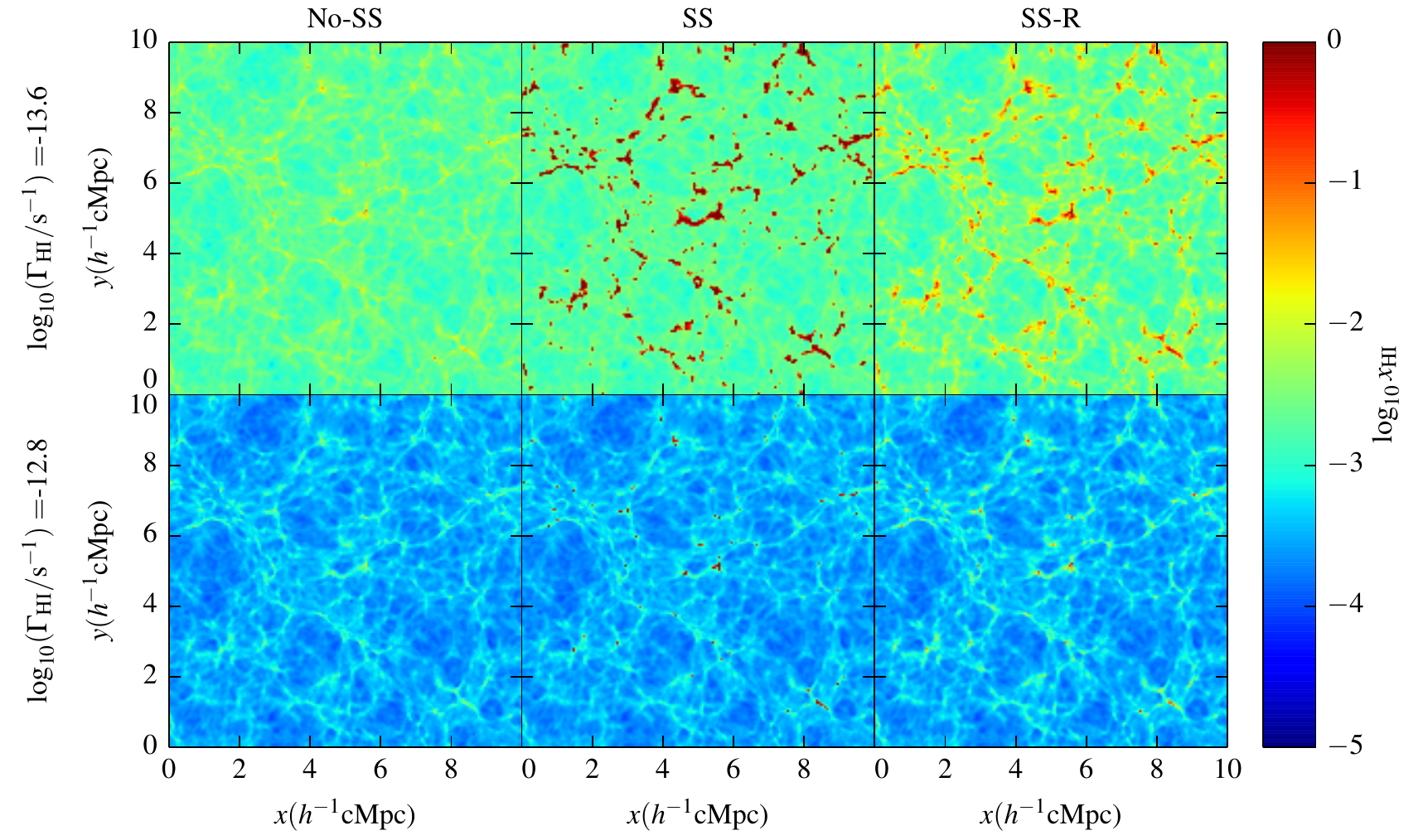}
\caption{The residual neutral hydrogen fraction in our small box at
  $z=7$. For the purpose of this comparison, we have assumed the small
  box lies entirely within an ionized region in our large box. Results
  are shown for different values of the background photoionization
  rate, $\bar{\Gamma}_{\rm HI}=-13.6$ (top) and $-12.8$ (bottom), and
  for the three different treatments of self-shielding we consider
  (from left to right, No-SS, SS and SS-R).}
\label{fig:plot_ionization_maps_smallbox}
\end{figure*}

\noindent
The spatial distribution of the neutral hydrogen fraction, $n_{\rm
  HI}/n_H$, obtained using these different models are compared in
\fig{fig:plot_ionization_maps_smallbox}.  The redshift chosen is $z =
7$ and results are shown for two background photoionization rates
$\log_{10} (\Gamma_{\rm HI} / \mathrm{s}^{-1}) = -12.8$ and $-13.6$.
The occurrence of fully neutral regions is almost negligible for the
No-SS case.  In the SS model the presence of overdense regions with
$x_{\rm HI} = 1$ is quite prominent.  In comparison, as discussed by
\citet{2014MNRAS.438.1820K} and \citet{2015MNRAS.446..566M} the effect
of self-shielding in the radiative transfer motivated SS-R model is
considerably reduced.  The volume-averaged neutral hydrogen fraction,
$\bar{x}^V_{\rm HI}$, for the three models is listed in Table
\ref{tab:xHI_vol}.

\begin{table}
\begin{tabular}{|l|c|c|}
\hline
Model & \multicolumn{2}{c}{$\bar{x}^V_{\rm HI}$} \\
\hline
& $\log_{10} (\Gamma_{\rm HI} / \mathrm{s}^{-1}) = -13.6$ & $\log_{10} (\Gamma_{\rm HI} / \mathrm{s}^{-1}) = -12.8$\\
\hline
No-SS & $2.1 \times 10^{-3}$ & $3.4 \times 10^{-4}$\\
SS & $3.0 \times 10^{-2}$ & $2.6 \times 10^{-3}$\\
SS-R & $6.5 \times 10^{-3}$ & $6.3 \times 10^{-4}$\\
\hline
\end{tabular}
\caption{The volume-averaged residual neutral hydrogen fraction at
  $z=7$ for the self-shielding models displayed in
  Fig.~\ref{fig:plot_ionization_maps_smallbox}.}
\label{tab:xHI_vol}
\end{table}

\section{Calibrating the reionization history in the hybrid simulation}
\label{sec:calibration}

\subsection{Constructing the hybrid simulation}

\begin{figure*}
\centering
\includegraphics[angle=0,width=0.95\textwidth]{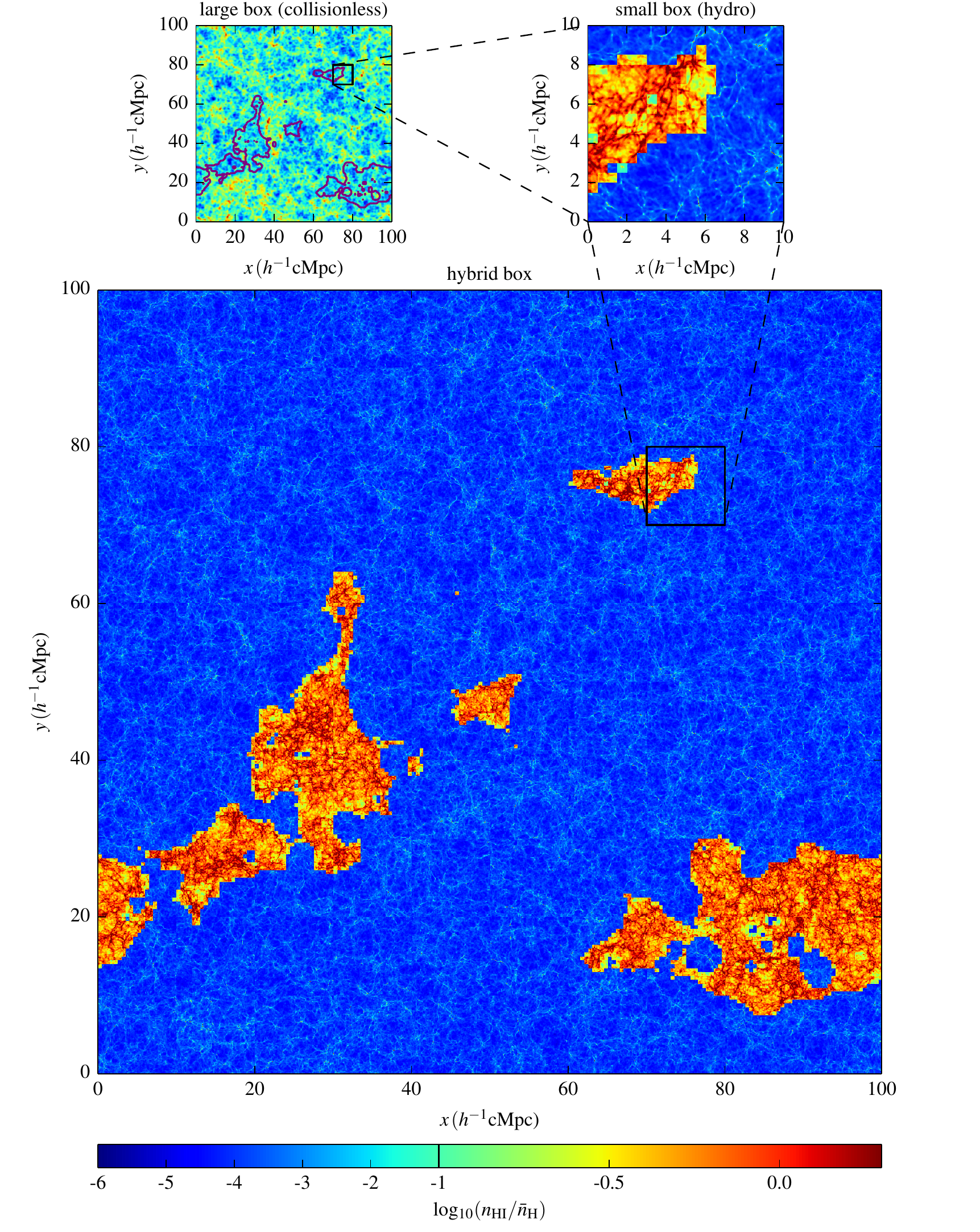}
\caption{Illustration of the construction of the high-resolution
  hybrid box including the large scale ionization field at $z=7$. The
  \textit{top left} panel shows the large scale density field obtained
  from the large box of size $100 \, h^{-1}$ Mpc, as well as the
  boundaries of the ionized regions obtained using the analytic
  model. The \textit{top right} panel displays the neutral hydrogen
  density in the small box with the map of the ionized regions from
  the large box applied to it. The \textit{lower} panel shows the
  neutral hydrogen density in the hybrid box, constructed by
  populating the volume of the large box with $10^3$ randomly shifted
  and rotated copies of the small box.}
\label{fig:plot_ionization_maps_big_smallbox}
\end{figure*}

We now turn to describe how we combine the small and large boxes in
our hybrid approach and calibrate the reionization history.  This
involves replicating the small box and superimposing the large scale
ionization map from the large box. The baryonic density field in the hybrid box is calculated using only the small box, while the peculiar velocity field is obtained by adding the large-scale modes from the large box to the velocity field of the small box (the details of this procedure are discussed in  Appendix \ref{app:velocity_field}).
 Figure \ref{fig:plot_ionization_maps_big_smallbox} illustrates this approach,
where we exploit the periodicity of the simulation and apply a random
translation and rotation to the particle positions and velocities in
each copy of the small box.  However, in order to correctly include
the \lya opacity of the intervening IGM in this hybrid volume, we must
calculate a photoionization rate consistent with the overall ionized
fraction of hydrogen in the large box.

There are still rather large uncertainties with regard to the
background photoionization rate during reionization. At the tail-end
of reionization at $5<z<6$, observational data are based on the
observed mean transmitted flux and the proximity effect in QSO
absorption spectra
\citep{2011MNRAS.412.1926W,2011MNRAS.412.2543C}. These measurements
utilise numerical hydrodynamical simulations which are very similar to
our small simulation box.  At $z>6$ the best constraints come from the
observed UV luminosity functions of high redshift galaxies
\citep[see][for recent results]{2013ApJ...768...71R}. Unfortunately
the latter probe the UV emissivity somewhat redwards of the
Lyman-limit and require considerable extrapolation in order to
estimate the ionizing emissivity.  As the recently updated UV
background model presented by HM2012 takes into account these and
other observations, we use this model as benchmark to calibrate the
reionization history of our simulation at $z > 6$. We now briefly
describe this calibration procedure below.  Further details are
discussed in Appendix \ref{app:ionization_field}.

\begin{figure*}
\centering
\includegraphics[angle=0,width=0.95\textwidth]{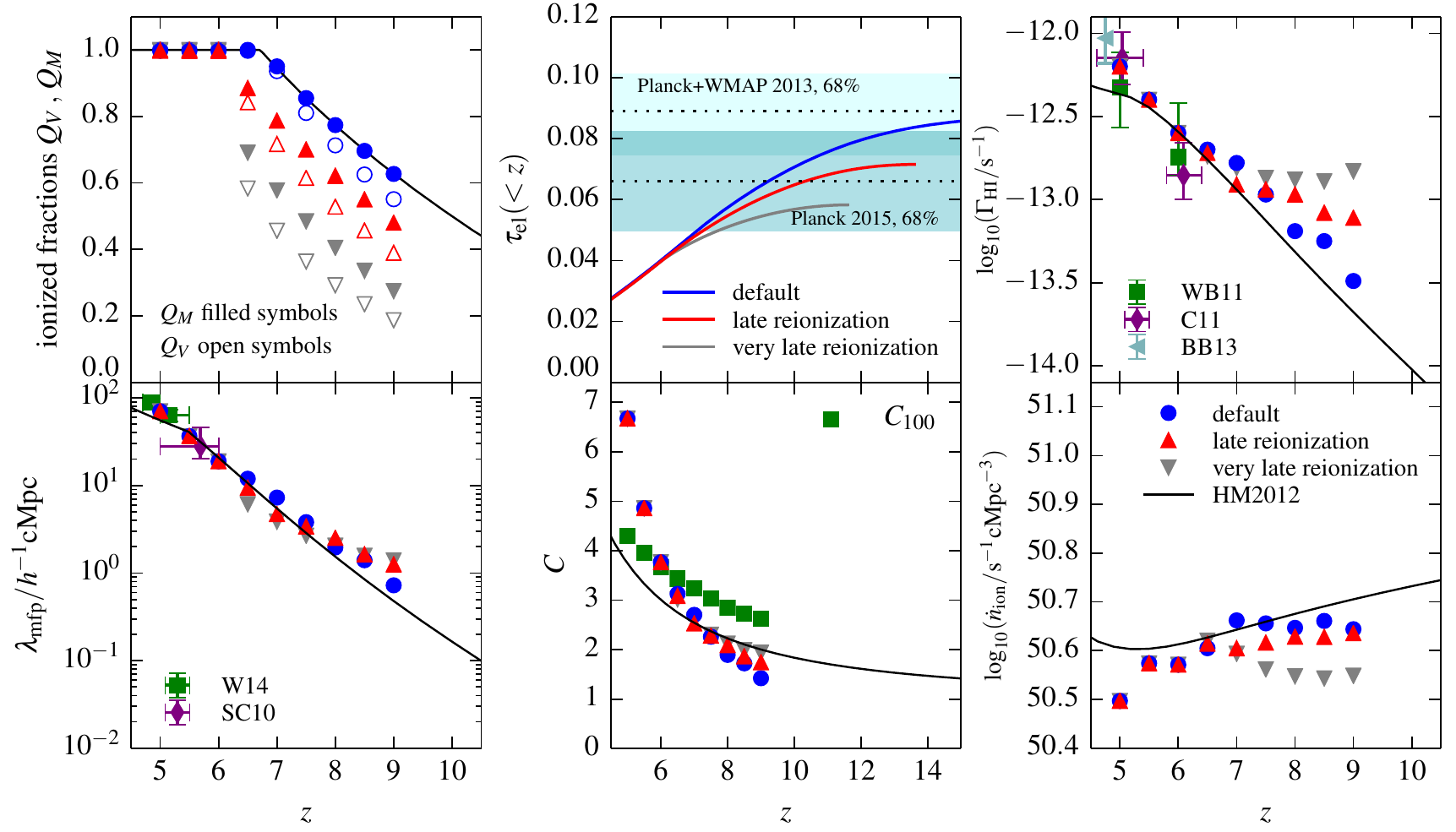}
\caption{Overview of the main properties of our reionization 
models.  Our default and late reionization models are shown in blue and red, respectively.  In light gray is shown 
a model, called the very late reionization model, where \lya emitters and ionized regions are assumed to be strongly correlated 
 (see  section  \ref{sec:lae_ew_dist} and Appendix \ref{sec:correlated}  for a detailed discussion). The \textit{black solid} curves show
  predictions of the HM2012 UV background model for reference. Our
  models are calibrated by fixing the evolution of the ionized mass
  fraction, $Q_{\rm M}(z)$. All other quantities are calculated
  self-consistently. For the default model, $Q_{\rm M}(z)$ was set
  equal to the ionized fraction in the HM2012 ``minimal reionization
  model''. The $Q_{\rm M}(z)$ chosen in our late reionization model is
  also shown in the \textit{upper left} panel, as well as the ionized volume fractions, $Q_{\rm V}(z)$. The \textit{upper
    middle} panel compares the optical depth due to Thomson
  scattering to the Planck 2013 and 2015 results
  \citep{2014A&A...571A..16P,2015arxiv150201589P}. 
The \textit{upper right} panel shows the average hydrogen photoionization rate within ionized regions obtained in our models.
  Observational
  constraints from \citet[][WB11]{2011MNRAS.412.1926W},
  \citet[][C11]{2011MNRAS.412.2543C} and \citet[][BB13]{2013MNRAS.436.1023B} are shown for reference. The
  \textit{lower left} panel displays the mean free path of ionizing
  photons at $1$ Ryd measured from our hybrid box. The data is from
  \citet[][W14]{2014MNRAS.445.1745W} and
  \citet[][SC10]{2010ApJ...721.1448S}. The \textit{lower middle} panel
  shows the clumping factor in the ionized regions. To allow a direct
  comparison to the clumping factor used in HM2012, we have also
  computed the clumping factor ${\cal C}_{100}$ of gas below an
  overdensity of 100. The \textit{lower right} panel shows the globally averaged
  comoving emission rates of ionizing photons that our models imply,
  compared to the rate predicted by the HM2012 UV background model.}
\label{fig:plot_HM12}
\end{figure*}

Firstly, in order to compute the Ly$\alpha$ optical depth in the
hybrid simulation at each redshift, we require the photoionization
rate, $\Gamma_{\rm HI}$, within the ionized bubbles.  We estimate
this self-consistently by the following iterative process.  We start
with a trial value of $\Gamma_{\rm HI}$, and for a given
self-shielding prescription we may then use this to estimate the
average mean free path for ionizing photons, $\lambda_{\rm mfp}$, and
the clumping factor of the gas, ${\cal C}$, within ionized bubbles
(see Appendix \ref{app:ionization_field} for details). The knowledge
of $\lambda_{\rm mfp}$ allows us to estimate the \emph{globally}
averaged comoving photon emissivity as
(e.g. \citealt{2012MNRAS.423..862K,2013MNRAS.436.1023B}) \be
\dot{n}_{\rm ion} = \f{\Gamma_{\rm HI}~Q_V}{(1+z)^2 \sigma_H
  \lambda_{\rm mfp}} \left(\f{\alpha_b + 3}{\alpha_s}\right),
\label{eq:Gamma_HI}
\ee where $\alpha_s$ is the spectral index of the ionizing sources (in
this case, the stars) at $\lambda < 912 {\rm\, \AA}$ and $\alpha_b$ is the
spectral index of the ionizing background. The factor $Q_V$, which is
the ionized fraction by volume, converts the photon emissivity in
ionized bubbles to the globally averaged value.  The quantity
$(\alpha_b + 3)/\alpha_s$ is estimated from the model of
\citet{2012ApJ...746..125H} by computing the ratio $\dot{n}_{\rm
  ion}\lambda_{\rm mfp}/(\Gamma_{\rm HI}~Q_V)$.

One can then estimate the time derivative of the \emph{mass}-averaged
ionization fraction, $Q_M$, from the equation
(e.g. \citealt{1999ApJ...514..648M}) \be \f{\der Q_M}{\der t} =
\f{\dot{n}_{\rm ion}}{n_H} - \f{Q_M}{t_{\rm rec}}.
\label{eq:dQdt}
\ee The recombination time scale $t_{\rm rec}$ is given by \be t_{\rm
  rec} = \f{1}{{\cal C}~\alpha_R~\chi~\bar{n}_H~(1+z)^3}, \ee where
$\alpha_R$ is the recombination rate coefficient for which we assume
the same value as HM2012.  We assume that helium is singly ionized in
H\,\textsc{ii} regions, so that $\chi = 1.08$ is the number of free
electrons per hydrogen nucleus.

We then repeat this procedure iteratively until the evolution of $Q_M$
is matched to that of HM2012 at $z<9$.  Thus, for each redshift, we choose a value of
ionizing efficiency $\zeta_{\rm eff}$ (see
Eq.~\ref{eq:zeta_eff_barrier}) and $\Gamma_{\rm HI}$ that reproduce the assumed $Q_M$ and $\der Q_M/\der t$ at that redshift. Note that this procedure breaks down in the
post-reionization era when $Q_M = 1$ and $\der Q_M/\der t = 0$. In
that case, we assume a value of $\Gamma_{\rm HI}$ which is consistent
with observations at $z \sim 5-6$. The Ly$\alpha$ optical depth can be
easily computed once $\Gamma_{\rm HI}$ is fixed.

\subsection{Reionization histories}

\begin{figure*}
\centering
\includegraphics[angle=0,width=0.95\textwidth]{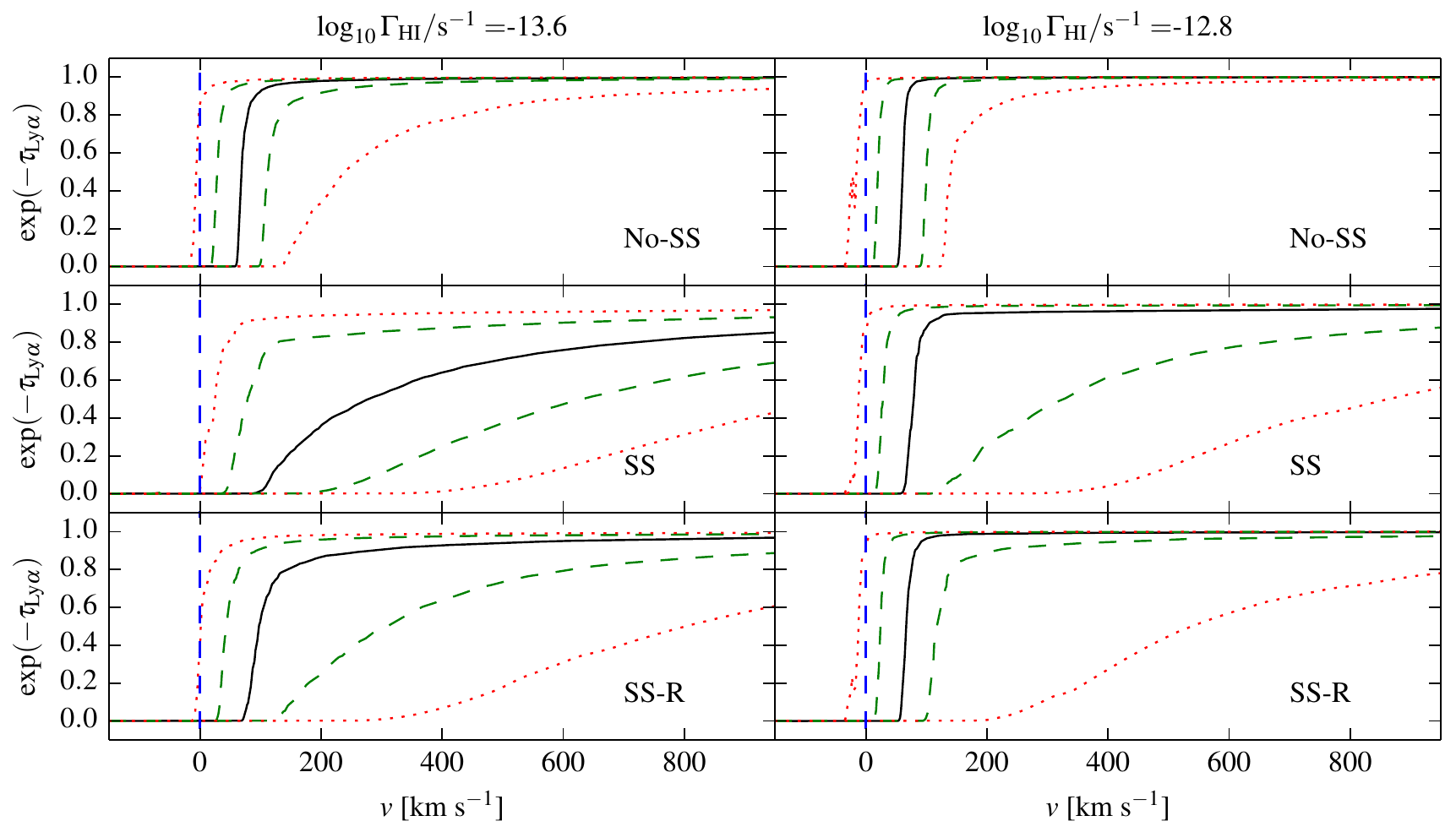}
\caption{The median and scatter of the Ly$\alpha$ transmission,
  $e^{-\tau_{\rm Ly\alpha}}$, along individual sightlines obtained
  from the small box at $z=7$. The results are shown for three
  different prescriptions for self-shielding and for two values of the
  background photoionization rate (assumed to be uniform). The dashed
  (green) and dotted (red) curves bound 68 and 95 per cent of the
  Ly$\alpha$ transmission around the median (black
  curves). 
}
\label{fig:plot_flux_stat_smallbox_003}
\end{figure*}

The iterative calibration of our model described above allows us to
fix one quantity, $Q_M(z)$.  All other quantities are then obtained
self-consistently from the hybrid simulation box. In this work we will
consider three reionization histories:

\begin{itemize}

\item \textbf{default reionization model:} $Q_M(z)$ is set equal to
  the ionized fraction in the HM2012 ``minimal reionization model'', as
  discussed above.\\

\item \textbf{late reionization model:} this model is motivated by the
  rather rapid evolution in ionized fraction between $z = 6$--$8$
  required to match the \lya emitter data. 
  We simply shift the
  $Q_M(z)$ such that the reionization is completed at $z = 6$ (as
  opposed to 6.7 in HM2012) keeping the value of $\der Q_M/\der z$
  unchanged.

\item \textbf{very late reionization model:} this model is introduced in order to match the \lya emitter data assuming \lya emitters and ionized regions are strongly correlated 
 (see  section  \ref{sec:lae_ew_dist} and Appendix \ref{sec:correlated}  for a detailed discussion). The ionized fraction in this case evolves more rapidly compared to the other two models. 

\end{itemize}

\noindent
The mass-averaged ionization fraction, $Q_M(z)$ for the three models is
shown in the upper left panel of Figure \ref{fig:plot_HM12}. The
figure also displays the evolution of several other quantities
describing the progress of reionization, along with a comparison to
observations and the HM2012 predictions.  By construction, we find
good agreement with the observed photoionization rates at $z=5-6$. We
also find reasonable agreement with the observed mean free path at
these redshifts. This is not surprising as these have been inferred
from the observational data using simulations very similar to our
small box. The ionizing emissivity in our default reionization model
differs very little ($\la 25\%$) from that assumed in HM2012.  The
difference is mainly due to a different clumping factor; HM2012 have
assumed a fixed density threshold of $\Delta_{\rm ss}=100$ for
self-shielding, while we have calculated it self-consistently.  The
clumping factor in our models therefore increases more rapidly with
decreasing redshift.  It is larger (smaller) than that in the HM2012
model at $z < 7 ~(z > 7)$.  The green squares in the lower middle
panel show that we get a very similar clumping factor as HM2012 if we
choose the same fixed self-shielding threshold of $\Delta_{\rm
  ss}=100$. The small difference is due to a somewhat later
reionization redshift in our small box ($z=15$) compared to that of
the simulations in \citet{2009MNRAS.394.1812P} ($z=19.5$) on which the
HM2012 prediction for the clumping factor is based.  Once  we choose
the time evolution of $Q_M$ to be same as in HM2012 in our default
reionization model, the values of $\Gamma_{\rm HI}$ and $\lambda_{\rm mfp}$
are   close to those   in HM2012. At high redshift the photoionization rate is slightly higher as the  photoionization rate within ionized regions is weighed inversely by the volume filling factor $Q_V$.

In our late reionization model the properties at $z \leq 6$ remain
identical to the default model. At higher redshift, there are moderate
changes in the quantities of interest as shown in
Fig.~\ref{fig:plot_HM12}. We find that in this model, the electron
scattering optical depth is reduced to $\tau_{\rm el} = 0.072$ as
opposed to $\tau_{\rm el} = 0.086$ in the default model.  The value of
$\tau_{\rm el}$ in the late reionization model is therefore lower than
the 2013 Planck constraints obtained from Planck temperature and WMAP
polarization data \citep{2014A&A...571A..16P}, but is fully consistent
with the latest Planck result $\tau_{\rm el} = 0.066 \pm 0.016$ \citep{2015arxiv150201589P}.

\section{The transmissivity of the \lya line}
\subsection{Damping wings redwards of the systemic redshift}

We now examine the \lya opacity arising from the IGM in our models.
To make contact with the earlier work of \citet{2013MNRAS.429.1695B},
we therefore first discuss the effects of the self-shielding
prescription on Ly$\alpha$ absorption spectra. For this, we need to
construct sightlines from the simulation box.  We use the following
method for calculating the absorption spectra: (i) we first extract
sightlines parallel to the box boundaries through the most massive
dark matter haloes in the simulation box. (ii) Each of these
sightlines is spliced with other randomly drawn sightlines in the box
to form a sightline of length $100 \, h^{-1}$ cMpc.  (iii) The
Ly$\alpha$ optical depth, $\tau_{{\rm Ly}\alpha}$, is estimated along
each line of sight given a value for $\Gamma_{\rm HI}$. We should
mention that we do not attempt to model the complexities arising from
radiative transfer within the host halo, hence as in
\citet{2013MNRAS.429.1695B} we ignore the contribution of any neutral
gas within 20 pkpc of the centre of the host halo.

The distribution of the transmitted fraction, ${\rm e}^{-\tau_{{\rm
      Ly}\alpha}}$, for 600 such sightlines is shown in
\fig{fig:plot_flux_stat_smallbox_003} for two different values of the
photoionization rates and for different self-shielding
prescriptions. The corresponding neutral volume fraction for these
models are given in Table~1.  In each panel, we show the median and 68
and 95 per cent bounds of the scatter around the median transmitted
fraction. Without self-shielding (top panel), the median value of
transmitted fraction recovers to nearly unity for rather small
velocity shifts ($\sim 100$ km s$^{-1}$) and the scatter is relatively
small.  As discussed in \citet{2013MNRAS.429.1695B}, in the SS model
(middle panel) the transition is significantly reduced out to much
larger velocity shifts due to the damping wing arising from optically
thick regions. In addition, there is a large scatter around the median
value once self-shielding is taken into account.  For the SS-R model,
the median value of the transmitted fraction is larger compared to
the SS model, although not as large as in the No-SS case.  However,
\emph{the scatter remains similar to that observed in the SS case.}
This suggests that the spatial distribution of the transmitted
fraction through the IGM will be highly variable when including the
\lya opacity in ionized regions -- we will investigate this in the
next section with maps of the spatial distribution of the \lya
emission line transmissivity.  Note also we have \emph{not} used any
information from the large box, i.e., the large scale patchiness in
the ionization field in this section.

\subsection{The large scale distribution of the \lya emission line transmissivity}

\begin{figure*}
\centering
\includegraphics[angle=0,height=0.94\textheight]{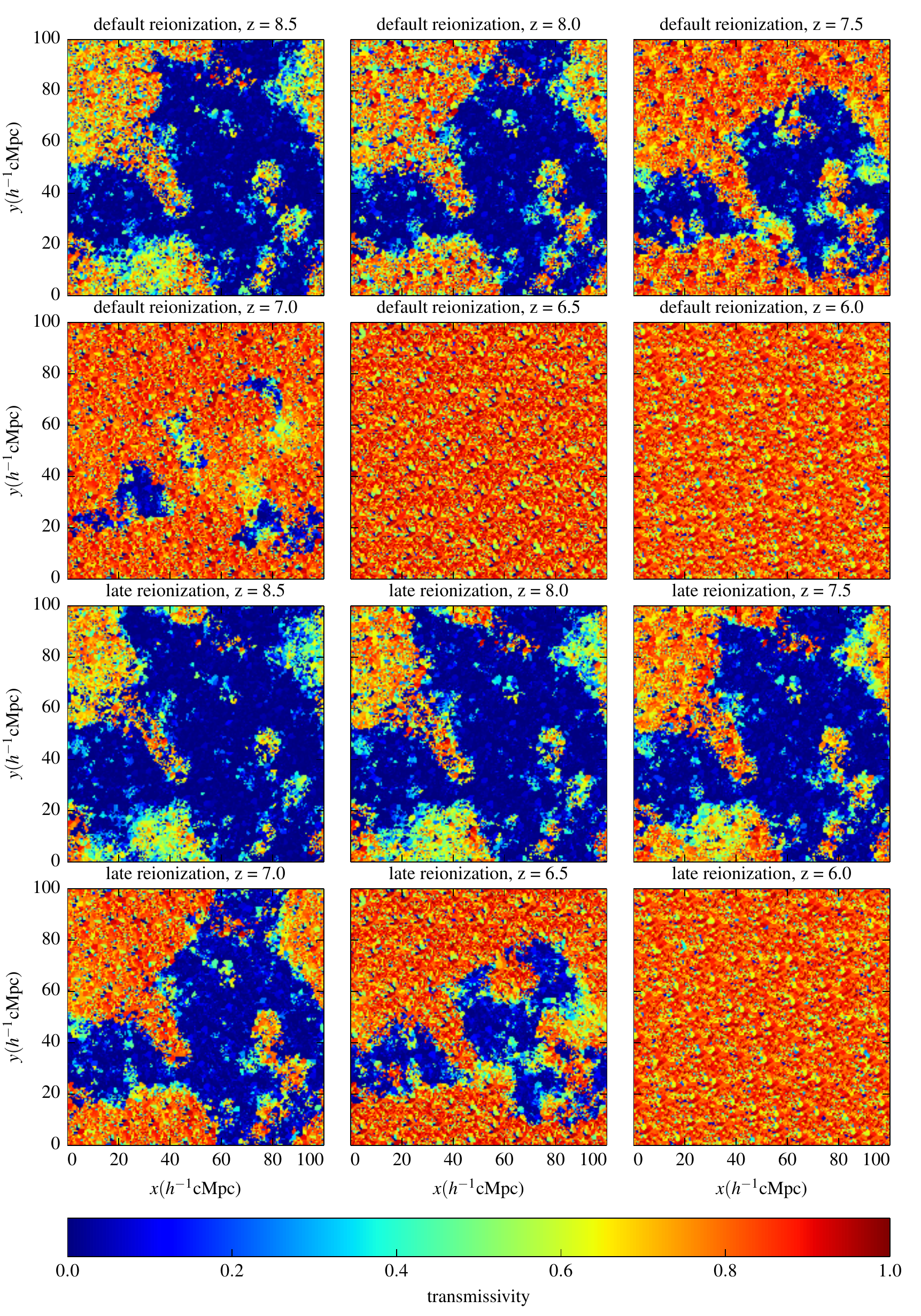}
\caption{Maps of the transmissivity for \lya emission lines
  at different redshifts. The two \textit{upper} rows show results for
  the default reionization model, while the two \textit{lower} rows
  show results for the late reionization model.  The intrinsic
  velocity shift is $\Delta v_{\rm int} = 100$ km s$^{-1}$, and the
  width of the Ly$\alpha$ profile is $\sigma = 88$ km s$^{-1}$. The
  self-shielding is implemented according to the SS-R model. The maps
  are coloured according to the transmissivity of the nearest emitter
  as seen in projection.}
\label{fig:plot_9Tmaps_halo_HM}
\end{figure*}

Using the spatial neutral hydrogen distribution constructed as
described in section \ref{sec:calibration}, we are now in a position
to investigate the large scale distribution of the transmissivity for
the dark matter haloes expected to host \lya emitters. An example of
this is shown in Figure \ref{fig:plot_9Tmaps_halo_HM} for both our
default and late reionization histories. We have identified the 250
most massive dark matter haloes in the small box and located the
positions of all their copies in a $10\,h^{-1} \mathrm{cMpc}$ thick
slab of the hybrid box.  As the slab is populated with 100 shifted and
rotated copies of the small box, this procedure yields 25000
sightlines towards these haloes for which we estimate the \lya line
transmissivity \be T = \f{\int_{\nu_{\rm min}}^{\nu_{\rm max}} \der
  \nu~J(\nu)~{\rm e}^{-\tau_{{\rm Ly}\alpha}(\nu)}}{\int_{\nu_{\rm
      min}}^{\nu_{\rm max}} \der \nu~J(\nu)}, \ee where $J(\nu)$ is
the Ly$\alpha$ line profile. In this work, we take the shape of the
profile to be Gaussian of width $\sigma$ with its centre shifted
redwards by an amount given by $\Delta v_{\rm int}$.  We also account
for the large-scale velocity fields from the large box (the details of
this procedure and its effects are discussed in Appendix
\ref{app:velocity_field}).

The evolution of the transmissivity along these 25000 sightlines are
shown as two-dimensional maps for a thin slice of our simulations in Figure \ref{fig:plot_9Tmaps_halo_HM}
for our two reionization models. Note that the maps do not  represent the  expected distribution of
transmissivity encountered by  an observationally selected sample of  Ly$\alpha$ emitters at that redshift, they show the value of $T$  that  a Ly$\alpha$ emitting galaxy  would experience if it were located in the slice of the simulation 
for which the transmissivity is plotted. 
The maps are coloured according to
the transmissivity of the nearest halo as seen in projection, where a
2D Voronoi tessellation is used for colouring the maps. The top two
rows are for the default model, while the bottom two are for the late
reionization model. For Figure \ref{fig:plot_9Tmaps_halo_HM} we have
assumed the \lya line to have a width $\sigma = 88$ km s$^{-1}$ and to be
shifted redward by an amount $\Delta v_{\rm int} = 100$ km s$^{-1}$, which at
$z \sim 6$ is similar to values recently inferred from the
CIII]$\lambda$1909 line \citep{2015MNRAS.450.1846S}. Note that this is
  a smaller velocity shift than the 200-400 km s$^{-1}$ considered by
  \citet{2015MNRAS.446..566M}.

\begin{figure*}
\centering
\includegraphics[angle=0,width=\textwidth]{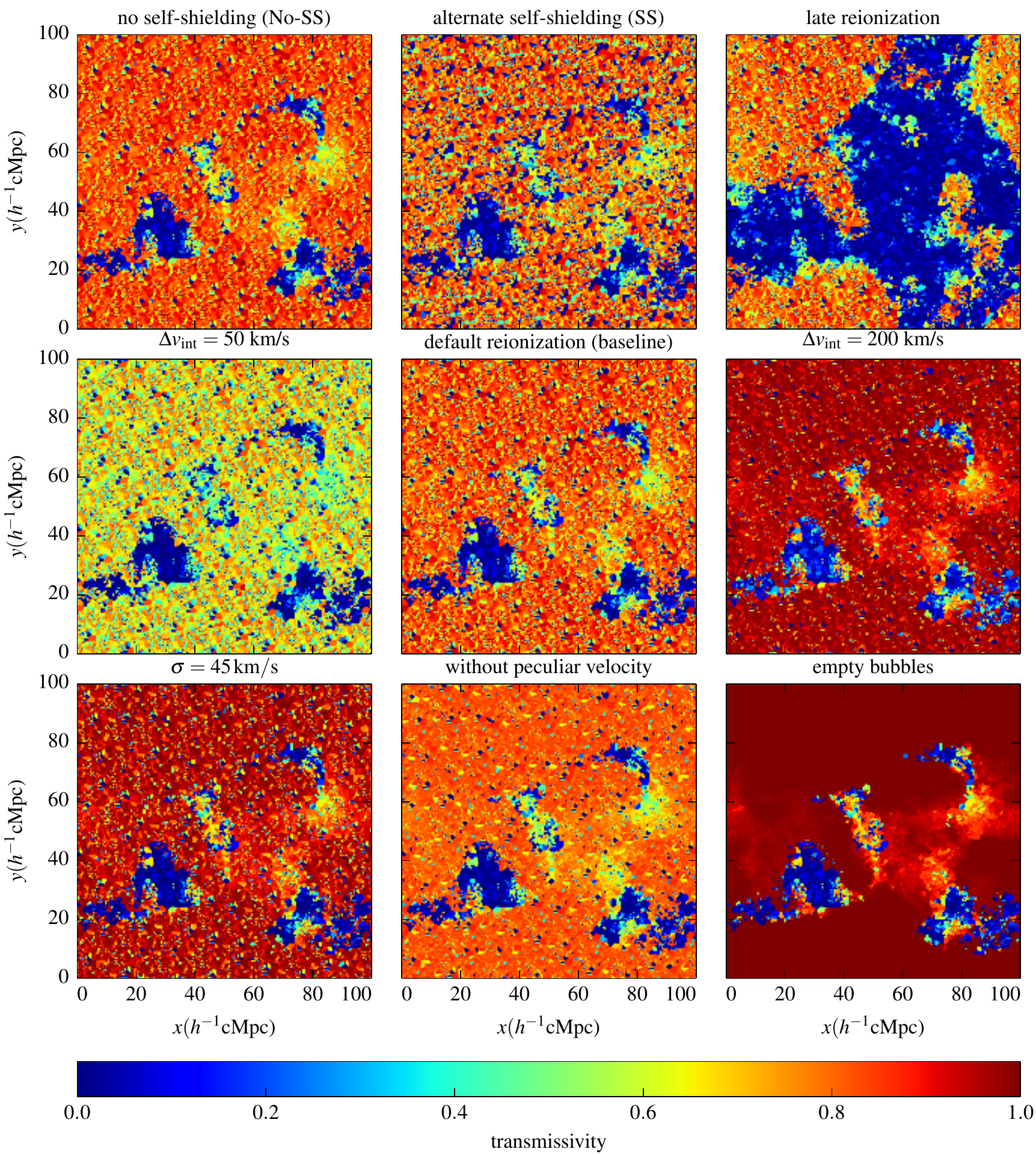}
\caption{Maps of the transmissivity for \lya emission lines
  for different model parameters. Results are shown at redshift $z =
  7$. The baseline model shown in the \textit{central} panel assumes
  the default reionization model, with $\Delta v_{\rm int} = 100$ km
  s$^{-1}$, $\sigma = 88$ km s$^{-1}$, the SS-R self-shielding model
  and peculiar velocities included (identical to the left panel in the
  second row of Fig.~\ref{fig:plot_9Tmaps_halo_HM}). The other panels
  show the effect of changing one of these assumptions at a
  time. Starting at the \textit{top left} and going clockwise the
  panels show the transmissivity without self-shielding, with the SS
  self-shielding model, with the late reionization model, for $\Delta
  v_{\rm int} = 200$ km s$^{-1}$, ignoring absorbers in ionized
  bubbles, neglecting the effect of the peculiar velocity, for $\sigma
  = 45$ km s$^{-1}$ and for $\Delta v_{\rm int} = 50$ km s$^{-1}$,
  respectively. The maps are coloured according to the transmissivity
  of the nearest emitter as seen in projection.}
\label{fig:plot_Tmaps_halo_parameters}
\end{figure*}

The mostly blue regions in Figure \ref{fig:plot_9Tmaps_halo_HM} are
not yet reionized and transmission is very low. In the already ionized
regions there is a large object-to-object scatter but as expected the
median transmissivity is still substantially smaller than unity.  The
transmissivity thereby depends strongly on the intrinsic redshift and
shape of the \lya emission line.  In Figure
\ref{fig:plot_Tmaps_halo_parameters}, we show the effect of various
parameters on the transmissivity maps at a representative redshift of
$z = 7$. The middle panel shows the default reionization model
assuming the SS-R self-shielding model with $\sigma = 88$ km s$^{-1}$ and
$\Delta v_{\rm int} = 100$ km s$^{-1}$. The left and middle panels in the top
row show how the modelling of the self-shielding affects the
transmissivity. As before the SS-R prediction falls between those
without self-shielding and the SS model.  In the middle row we compare
transmissivity maps for different values of the redward shift $\Delta
v_{\rm int}$ of the emission line. With decreasing intrinsic redshift
of the \lya emission the transmissivity due to neutral hydrogen in the
IGM in front of the emitters rapidly decreases. The bottom left panel
shows the effect of reducing the line-width from $\sigma=88$ km s$^{-1}$ to
$\sigma=45$ km s$^{-1}$ as  a reduced intrinsic velocity shift may be correlated  with a reduced line width due to the
resonant nature  of Lyman-alpha scattering. The average transmissivity
increases  somewhat for the reduced line width.
For the somewhat larger line widths often observed  at
lower redshift \citep[e.g.,][]{2006PASJ...58..313S} the average
transmissivity would be  somewhat smaller than in our default model.  The bottom middle panel shows the effect of
peculiar velocities. As discussed in more detail in Appendix
\ref{app:velocity_field}, the effect of peculiar velocities is rather
weak as the relevant gradients in peculiar velocity between host
haloes and absorption from the surrounding IGM are rather
small. Finally in the bottom right panel we show the case where the
ionization field is constructed only from the large box, i.e., the
ionized bubbles contain no neutral or partially ionized gas at all. As
expected, in this case the transmission is only reduced behind not yet
ionized regions.

The spatial patterns in transmissivity maps discussed above are due to
a convolution of the underlying ionization field and the peculiar
velocities of the \lya host haloes and the intervening IGM. This
spatial pattern will be further modified by spatial fluctuations in
the photoionization rate within ionized regions, which our
simulations do not yet model as we assume a single mean free path.
Likely uncorrelated object-to object variation of the intrinsic
spectral profiles of the \lya emission and the spatial distribution of
any \lya absorbing dust will also modify this picture further. These
transmissivity maps should nevertheless give a qualitative feel for
the effect of the inclusion of the intervening \lya opacity due to
residual hydrogen in optically thick regions in otherwise already
ionized regions. 
Note that  the area of the GOODS-N and GOODS-S fields are smaller
than the projected  size of our large simulation box and that the transmissivity maps in Figures \ref{fig:plot_9Tmaps_halo_HM} and \ref{fig:plot_Tmaps_halo_parameters} suggest
that there may be significant field to field variations for surveys of such  angular area.
More meaningful quantitative modelling of the clustering
of \lya emitters
(e.g. \citealt{2007MNRAS.381...75M,2013MNRAS.428.1366J}) which
incorporates this will require more detailed modelling tailored to
specific surveys, which we leave to future work.

\begin{figure*}
\centering
\includegraphics[angle=0,width=0.95\textwidth]{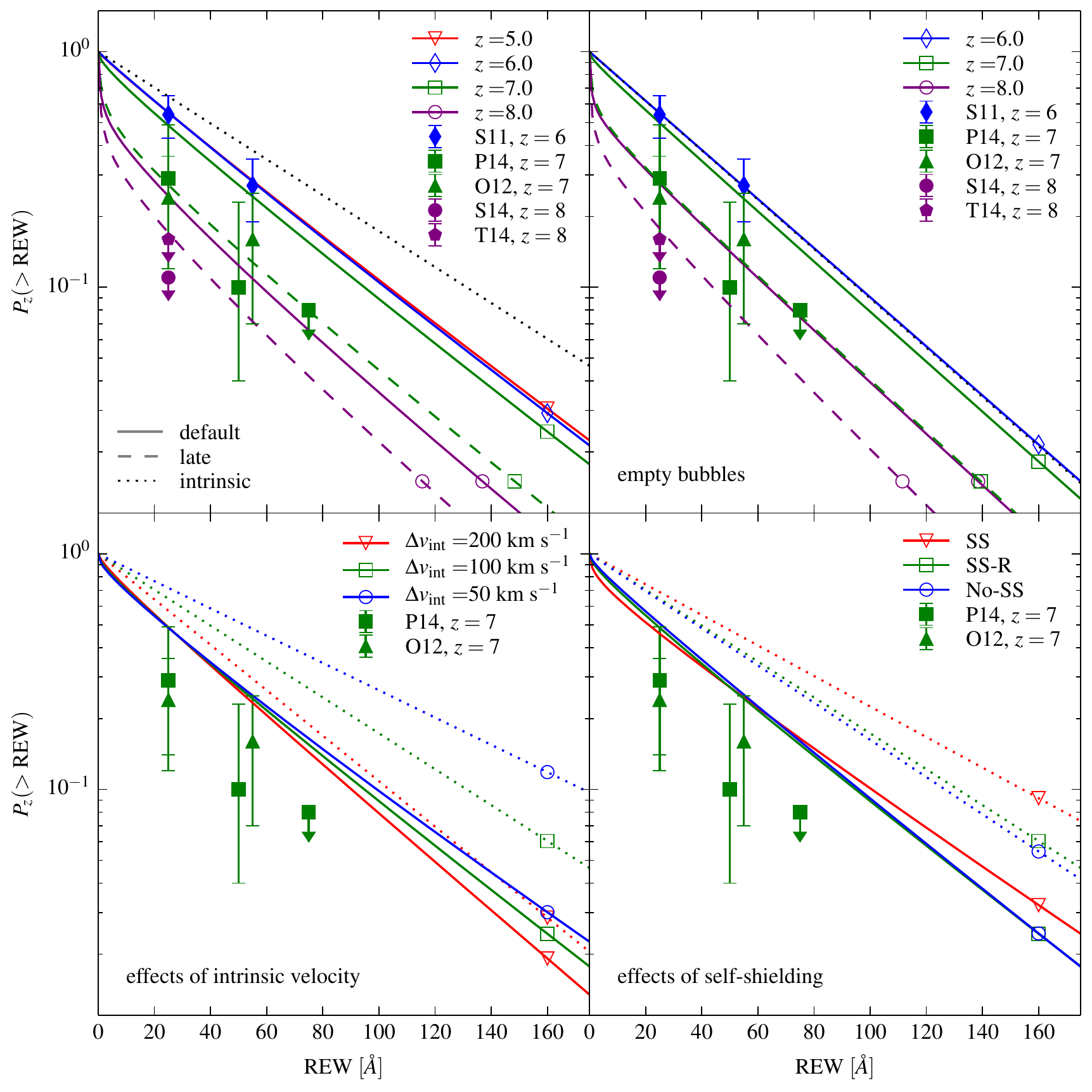}
\caption{Cumulative probability distributions for the rest-frame
  \lya equivalent width (REW).  The dotted curves represent
  the intrinsic REW distribution that reproduces the observed $z=6$
  distribution for that parameter set. The points with error-bars
  represent the observed data for faint UV galaxies with
  $\mbox{M}_{\rm {UV}} > -20.25$
  (\citealt{2011ApJ...728L...2S} - S11, composite data compiled by \citealt{2012ApJ...744...83O} - O12, \citealt{2014ApJ...793..113P} - P14, \citealt{2014ApJ...795...20S} - S14, \citealt{2014ApJ...794....5T} - T14). The
  \textit{top left} panel shows results for the default (solid curves)
  and late (dashed curves) reionization models assuming $\Delta v_{\rm
    int} = 100$ km s$^{-1}$ , $\sigma = 88$ km s$^{-1}$ and SS-R
  self-shielding. The \textit{top right} panel shows results obtained
  ignoring all absorbers in ionized bubbles. All other parameters are
  unchanged. The \textit{lower left} and \textit{right} panels show
  the effects of changing $\Delta v_{\rm int}$ and the self-shielding
  prescription, respectively. Line markers have been added near the
  lower right corner of each panel to better facilitate identification
  of the different colours.}
\label{fig:plot_lae_ewdist}
\end{figure*}

\subsection{The evolution of the observed \lya equivalent width distribution and the  rapid demise of \lya emitters at $z>6.5$}
\label{sec:lae_ew_dist}

Having calculated the \lya transmissivity due to neutral hydrogen for
a large number of \lya host haloes we are now in a position to
reassess the effect of an inhomogeneously reionized intervening IGM on
the evolution of \lya emitters. Previous studies have done this by
comparing to the observed $z=6$ equivalent width  distribution of
\lya emitters assuming it is not yet affected by the intervening IGM
(e.g. \citealt{2011MNRAS.414.2139D}) which is probably a good assumption for large velocity shifts $\Delta v_{\rm int} \sim 300-400$ km s$^{-1}$.  However, as is apparent from Figure
\ref{fig:plot_9Tmaps_halo_HM}, the intervening IGM will affect the
visibility also at $z\la 6$ for lower values of $\Delta v_{\rm int} \sim 100$ km s$^{-1}$.
Hence, for each model under
consideration, we prefer to instead use the unabsorbed rest-frame
\lya equivalent width (REW) distribution $P_{\rm
  int}(>\mbox{REW})$ that would be required to produce the observed $z
= 6$ distribution $P_{z=6}(>\mbox{REW})$ from
\citet{2011ApJ...728L...2S}.\footnote{We find that a simple
  exponential function $P_{z=6}(>\mbox{REW}) =
  \exp[-\mbox{REW}/\mbox{REW}_c]$ with $\mbox{REW}_c = 41.5 {\rm \AA}$
  fits the data points of \citet{2011ApJ...728L...2S} quite
  nicely. The value of $\mbox{REW}_c$ is somewhat smaller than used by
  \citet{2011MNRAS.414.2139D} based on the data from
  \citet{2010MNRAS.408.1628S}.}  As in \citet{2015MNRAS.446..566M} we
consider the inferred REW distribution of observed UV-faint galaxies
with M$_{\rm UV} > -20.25$ for which numbers are sufficient for a
reasonably robust determination of REW upper limits.

Following \citet{2011MNRAS.414.2139D}, the REW distribution at any redshift $z$ is calculated as
\be
P_z(>\mbox{REW}) = \int_0^1 \der T~P_{T,z}(T)~P_{\rm
  int}(>\mbox{REW}/T), 
\ee 
where $P_{T,z}(T)$ is the distribution of
Ly$\alpha$ transmissivity $T$. Note that we have assumed that the
unabsorbed distribution $P_{\rm int}(>\mbox{REW})$ does not evolve
with redshift and that  there is no spatial correlation  between the  positions of potential Ly$\alpha$ emitters  and ionized bubbles. As it is not known how strongly \lya emitters are biased at $z > 6$ there is no obvious way 
of accounting for such a correlation. We will discuss  later our third model where the positions of the \lya emitters are assumed to be strongly correlated with the ionized regions.
The resulting distributions are shown in
\fig{fig:plot_lae_ewdist}. In each panel, the dotted curves show the
unabsorbed REW distribution that was chosen to reproduce the observed
$z=6$ REW distribution \citep{2011ApJ...728L...2S} for each model. The
figure, thus, highlights the redshift evolution of the REW
distribution. We also show data points at $z = 6$
\citep{2011ApJ...728L...2S}, $z=7$ \citep{2012ApJ...744...83O,2014ApJ...793..113P} and the
upper limit at $z = 8$ \citep{2014ApJ...795...20S,2014ApJ...794....5T}. The top-left
panel shows how the REW distribution evolves with redshift for the
default and late reionization models with intrinsic velocity shift
$\Delta v_{\rm int} = 100$ km s$^{-1}$ and with our standard
self-shielding prescription SS-R. The top-right panel shows the
evolution when the ionized bubbles are assumed to contain no neutral
hydrogen at all. Our default model is clearly unable to explain the
rapid demise of the \lya emitters between $z = 6$ and $z = 8$ . The
predicted REW distribution does not fall fast enough. Interestingly,
our late reionization model matches the observed rapid fall at $z =
7$, and is close to (but still larger than) the upper limit at $z= 8$.
The bottom-left panel of Figure \ref{fig:plot_lae_ewdist} shows the
effects of using a different (but also redshift-independent) redwards
shift $\Delta v_{\rm int}$, while the bottom-right panel shows the
effects of the self-shielding prescription at $z = 7$.  Note that once
the distribution is normalized to the observed points at $z=6$ and the
parameters and $Q_{\rm M}(z)$ are kept fixed, none of these choices make much of a
difference to the predicted distribution at $z=7$.

We confirm the findings of \citet{2014MNRAS.438.1820K} and
\citet{2015MNRAS.446..566M} that the effect of self-shielding of neutral
hydrogen in LLSs is significantly reduced with the SS-R prescription
compared to the SS prescription and is overall rather weak unless the
photoionization rate is significantly lower than predicted by our
models.  As already discussed, however, we have probably
underestimated the effect of self-shielding neutral gas for a variety
of reasons: the missing large scale power in the matter distribution,
the missing correlation of ionization and density fields on large
scales and the neglect of fluctuations in the UV background amplitude
and mean free path (e.g. \citet{2015MNRAS.447.3402B}).

\begin{figure*}
\centering
\includegraphics[angle=0,width=0.9\textwidth]{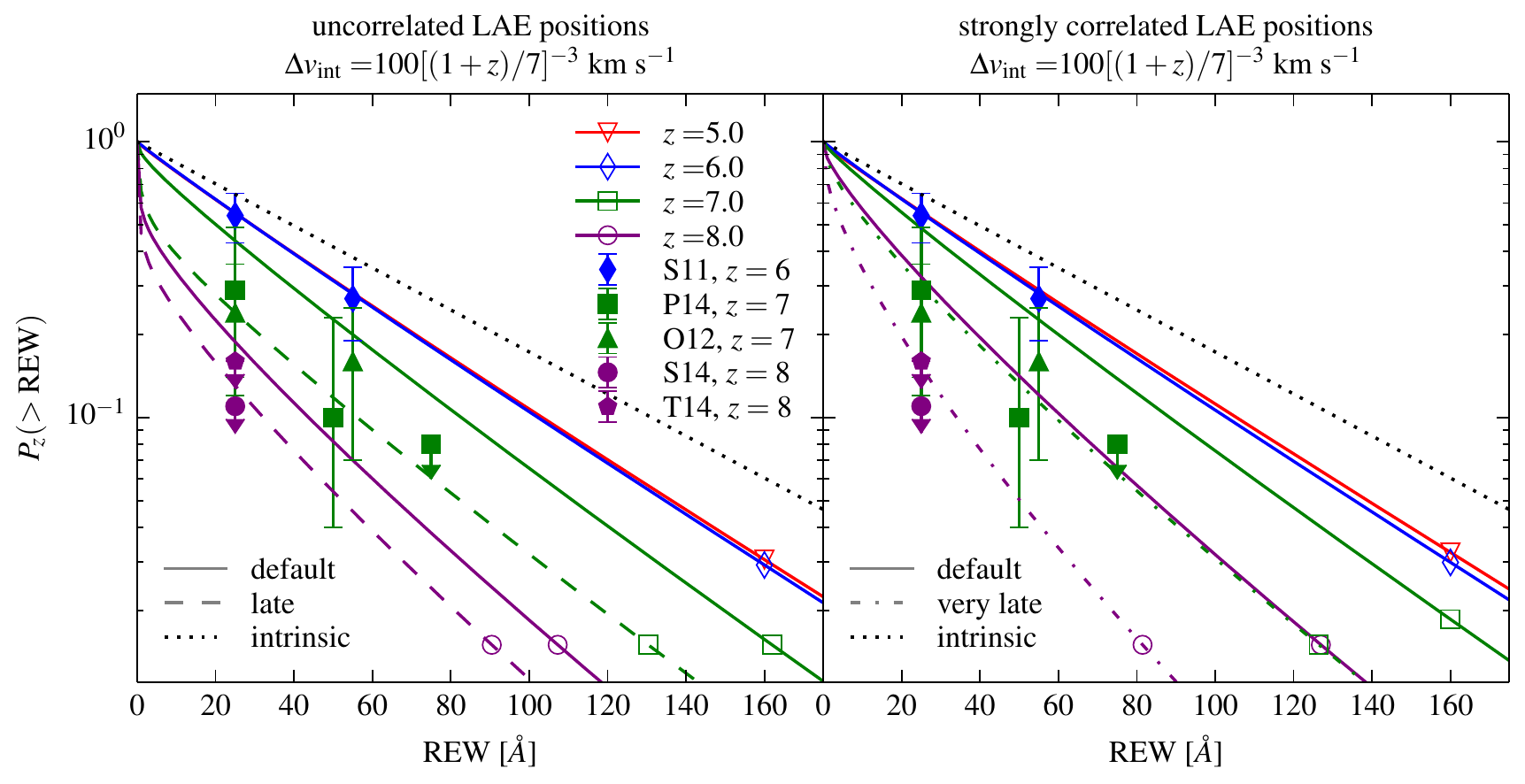}
\caption{Evolution of the REW distribution when the intrinsic redward
  shift $\Delta v_{\rm int}$ of the Ly$\alpha$ line is assumed to
  evolve with redshift as $\Delta v_{\rm int} = 100 \mbox{km s$^{-1}$}
  \left[(1+z)/7\right]^{-3}$. The left hand panel is for the case where the correlation between \lya emitters and ionized regions is ignored, while the right hand panel is for the case where the correlation is accounted for. All other parameter are the same as in
  the upper left panel of Fig.~\ref{fig:plot_lae_ewdist}. The points with error bars
  represent the observed data for faint UV galaxies with
  $\mbox{M}_{\rm {UV}} > -20.25$
  (\citealt{2011ApJ...728L...2S} - S11, composite data compiled by \citealt{2012ApJ...744...83O} - O12, \citealt{2014ApJ...793..113P} - P14, \citealt{2014ApJ...795...20S} - S14, \citealt{2014ApJ...794....5T} - T14).
}
\label{fig:plot_lae_ewdist_vwind_evol}
\end{figure*}

\begin{figure*}
\centering
\includegraphics[angle=0,width=0.9\textwidth]{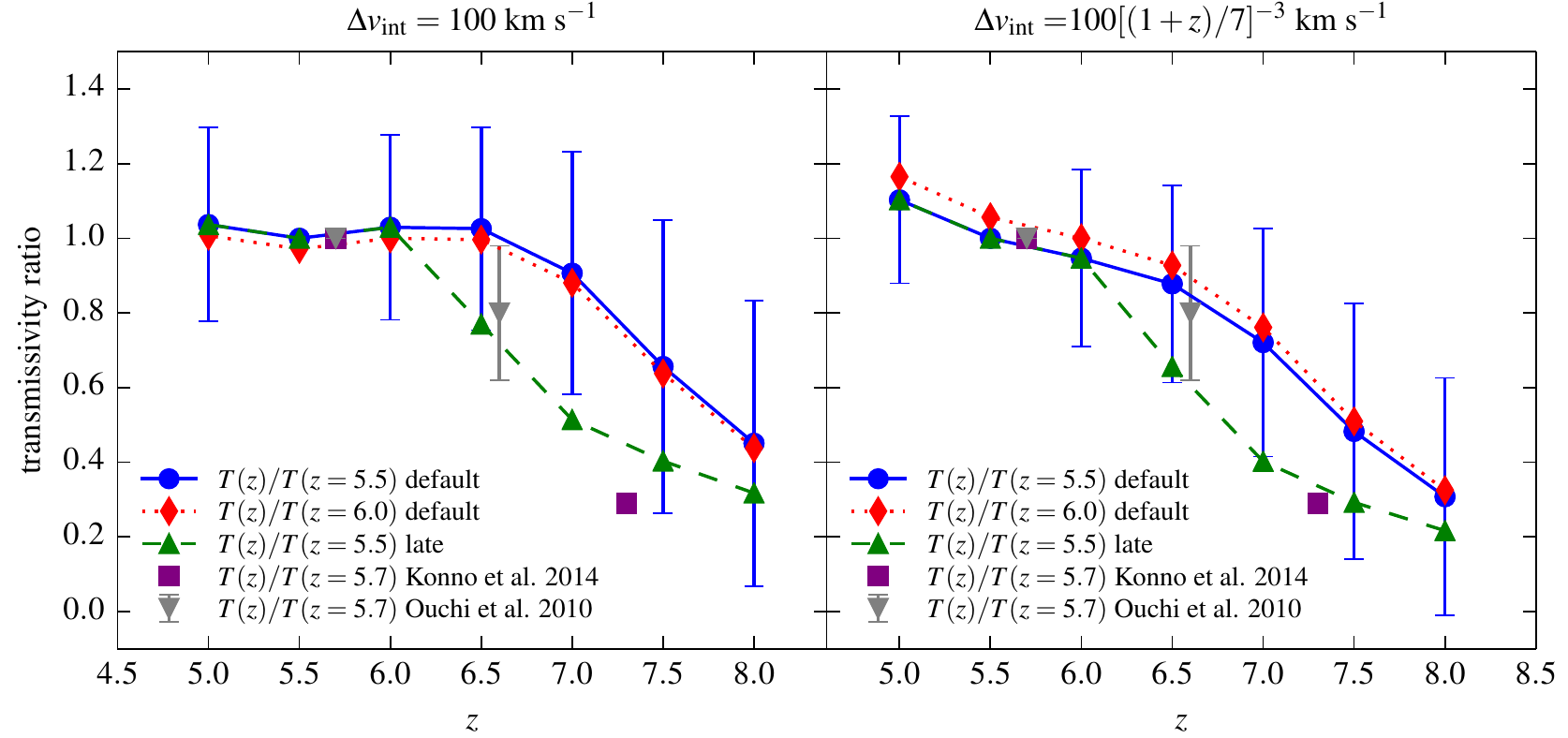}
\caption{Evolution of the mean \lya transmissivity of the
  IGM for the SS-R model, normalized to that at a particular redshift. The redward shift
  $\Delta v_{\rm int}$ of the Ly$\alpha$ line is $100 \mbox{km
    s$^{-1}$}$ in the \textit{left} panel and assumed to evolve with
  redshift as $\Delta v_{\rm int} = 100 \mbox{km s$^{-1}$}
  \left[(1+z)/7\right]^{-3}$ in the \textit{right} panel. The
  observational constraints are from \citet{2010ApJ...723..869O} and \citet{2014ApJ...797...16K}.  The
  error bars on the simulation data display the 68 per cent scatter
  for the 25000 sight-lines considered.}
\label{fig:plot_lae_tmean_vwind_evol}
\end{figure*}

In Figure \ref{fig:plot_Tmaps_halo_parameters} we had seen that the
transmissivity depends rather strongly on the intrinsic velocity
shift, $\Delta v_{\rm int}$, of the \lya emission \citep[see  also][]{2007MNRAS.377.1175D,2011ApJ...728...52L}.  The recent
measurement of the velocity shift of \lya emitters at $z\sim 6-7$ of
about 100 km s$^{-1}$ \citep{2015MNRAS.450.1846S} is significantly
smaller than the 300-400 km s$^{-1}$ typically observed in LBGs at
$z\sim 2$--$3$ \citep{2010ApJ...717..289S}. It is not yet clear what
causes this difference, but a plausible explanation is a decreasing
neutral hydrogen column density \citep{2011MNRAS.416.1723B} as the
interstellar medium in high-redshift galaxies becomes more ionized.
This may also explain the postulated increase in the escape fraction
of hydrogen ionizing photons with increasing redshift
\citep{2012MNRAS.423..862K,2013ApJ...768...71R,2013MNRAS.431.2826F,2013MNRAS.436.1023B}.
A rapidly decreasing $\Delta v_{\rm int}$ at $z>6$ would contribute
significantly to a decrease of the \lya transmissivity and may be more
plausible than the suggested rapid evolution of the \lya equivalent
width due to dust evolution \citep{2012MNRAS.421.2568D}. Note that smaller intrinsic velocity shifts  also enhance the effect of the self-shielding on the transmissivity.

To explore this possibility, we have also investigated a model were
the shift of the \lya line, $\Delta v_{\rm int}$, is evolving with
redshift, \be \Delta v_{\rm int} = 100 \mbox{km s$^{-1}$}
\left(\f{1+z}{7}\right)^{-3},
\label{eq:Delta_v_evol}
\ee which corresponds to a increase of about a factor two between
$z=8$ and $z=6$ from 50 to 100 km s$^{-1}$.  The results are shown in
the left hand panel of
\fig{fig:plot_lae_ewdist_vwind_evol} where the solid lines are for the
default model and the dashed lines are for the late reionization
model. The default model combined with such an evolution of $\Delta
v_{\rm int}$ still fails to match the rapid drop of the $z=7$ data
points.  However, the late reionization model with a neutral fraction
of $\sim$ 30 (50) per cent at $z=7$ ($z=8$) is now consistent with the
observed REW distribution.

In order to gauge the effect of  possible  correlations between the positions of the \lya emitting galaxies and the ionized regions, we have also studied a rather extreme  model  where we place the \lya emitters at the locations 
of the massive (i.e., $> 10^{10}$ M$_{\odot}$) haloes in the large box. Since sightlines towards  haloes in the large box would not account for the expected larger (than average) number of self-shielded regions around the emitters, we choose  the sightline toward the  halo in the small box which is nearest to the massive halo in the large box. 
Further details of how we choose the haloes are  given in Appendix \ref{sec:correlated}. To match the \lya emitter data using this approach, we find a suitable $Q_M(z)$ by trial and error
  such that the reionization is completed at $z = 6$ (as
  opposed to 6.7 in HM2012) and evolves rapidly at $z > 6$. Note that the evolution of $Q_M(z)$ in this model is considerably more rapid than in the  late reionization model discussed earlier.  The properties of this model, which we call the {\bf very late reionization} model, are shown as light gray points and  curves  in Figure~\ref{fig:plot_HM12}. The resulting REW distribution shown in the right hand panel of Figure~\ref{fig:plot_lae_ewdist_vwind_evol} is again reasonably consistent with the data. Placing the \lya emitters in haloes close to massive haloes in the large simulation box results in a  very strong (probably unrealistically strong) correlation between the location of the ionized bubbles and that of the \lya emitters  
for the  simple collapse fraction model employed here   for calculating the location of ionized bubbles.  
The smaller ionized mass fractions  required for the same reduction  in transmissivity should  thus 
probably better be considered as lower  limits.

Finally, an alternative way to probe the Ly$\alpha$ transmissivity of
the IGM is through the evolution of the Ly$\alpha$ luminosity
function, corrected for changes arising from the evolution of the star
formation rate density.  Using the ultra-deep Subaru narrowband
imaging survey for \lya emitters at $z = 6.6$ in SXDS field, \citet{2010ApJ...723..869O} estimate the ratio of \lya
transmissivity at $z = 6.6$ and $z=5.7$ to be $T(z=6.6)/T(z=5.7) =
0.80 \pm 0.18$. Similarly, using a  survey at $z= 7.3$ in SXDS and COSMOS
fields, \citet{2014ApJ...797...16K} estimate 
$T(z=7.3)/T(z=5.7) = 0.29$. In order to compare these results with our model predictions, we
plot the evolution of the mean transmissivity as obtained from the
maps shown in Figures \ref{fig:plot_9Tmaps_halo_HM} and
\ref{fig:plot_Tmaps_halo_parameters} in Figure
\ref{fig:plot_lae_tmean_vwind_evol}. The left panel assumes that
$\Delta v_{\rm int}$ = 100 km s$^{-1}$, while the right hand side is
for the case when the intrinsic velocity shift evolves according to
equation \eqref{eq:Delta_v_evol}. The evolution of the transmissivity
in our default model is again not rapid enough to account for observed
evolution when $\Delta v_{\rm int}$ does not evolve, but the expected
scatter is large.  The error bars (1 $\sigma$) show the object to
object variation predicted by our model. As expected, the late
reionization model is in better agreement with the data and matches
rather well if $\Delta v_{\rm int}$ decreases at high redshift. The results for the very late reionization model, accounting for the correlation of the location of \lya emitters and ionized regions,  are again very similar  to  that of the late model.

\section{Conclusions}

We have combined high resolution hydrodynamical simulations with an
intermediate resolution collisionless, dark matter only simulation and
an analytical model for the growth of ionized regions to estimate the
large scale distribution and redshift evolution of the visibility of
\lya emission in high-redshift galaxies.  We have carefully calibrated
the growth of ionized regions to that expected for the evolution of
the UV-background model of HM2012, and included the \lya opacity of
intervening, optically thick absorption systems within ionized
regions.  

The rapid evolution of the ionized volume fraction, mean-free path and
ionization rate of hydrogen at the tail end of reionization results in
a rapidly evolving and strongly spatially variable transmissivity for
the \lya emission line.  The \lya transmissivity is thus very
sensitive not only to the ionized volume fraction and amplitude of the
local ionizing background, but also the relative velocity shift of the
\lya emission.  Our default reionization model predicts an
evolution of the \lya REW distribution somewhat slower than observed,
suggesting that additional factors contribute to the observed rapid
demise of \lya emission with increasing redshift at $z>6$. The
predicted evolution is also somewhat slower than that found in
\citet{2013MNRAS.429.1695B}, mainly because of our improved modelling
of the self-shielding of neutral hydrogen based on the results of the
radiative transfer calculations by \citet{2013MNRAS.430.2427R}.

We have furthermore identified the observed apparent decrease of the
redwards shift of the intrinsic \lya emission with increasing redshift
\citep{2015MNRAS.450.1846S} as an important factor which may
contribute significantly to the evolution of the \lya emission at
$z>6$. Interestingly such a decreasing redwards shift may be
physically linked to the postulated increase in escape fraction of
ionizing radiation
\citep{2012MNRAS.423..862K,2013ApJ...768...71R,2013MNRAS.431.2826F,2013MNRAS.436.1023B}
and the expected lower neutral hydrogen column densities in
high redshift galaxies at $z>6$.  

A model where reionization completes
somewhat later and perhaps also more rapidly than assumed in the
HM2012 UV background model, and where the intrinsic velocity shift of
the \lya emitters increases from 50 to 100 km s$^{-1}$ between $z=8$
and $z=6$, matches the observed rapid decrease of the observed \lya
emission well. Based on the latest Planck results \citep{2015arxiv150201589P}, such a late
reionization is no longer disfavoured by cosmic microwave background
constraints. Accounting for the correlation of the location of ionized regions and \lya
emitters  further strengthens this conclusion.

Forthcoming wide angle \lya surveys aimed at at a
better characterization of the large scale clustering properties of
\lya emitters at $z\sim 6-8$, together with further improved modelling
of the spatial distribution of the \lya emission line transmissivity of
the kind we have presented here, should provide robust constraints on
the timing and patchiness of the reionization of hydrogen. 

\section*{Acknowledgements}

Support by ERC Advanced Grant 320596 ``The Emergence of Structure
during the epoch of Reionization" is gratefully acknowledged.  The
simulations used in this work were performed on the Darwin and Cosmos
supercomputers at the University of Cambridge.  Part of the computing
time was awarded through STFCs DiRAC initiative. JSB acknowledges the
support of a Royal Society University Research Fellowship. We thank Richard Ellis, Brant Robertson and Mark Dijkstra for helpful comments.

\bibliography{tirth14_17}

\appendix
\section{Method for generating ionization fields}
\label{app:ionization_field}

We outline here a more detailed description of the method we use to
generate the ionization field for our hybrid simulations. The steps
are as follows:

\subsection {Large scale ionization field} 

The first step is to generate the ionization field in the large
box. We identify the haloes using a friends-of-friends group finder
and assign emissivities proportional to the halo mass. These are
subsequently used for generating the ionization field. A given
location $\mathbf{x}$ in the simulation box is assumed to be ionized
if, within a spherical region of radius $R$ around it, the condition
\be \zeta f_{\rm coll}(\mathbf{x}, R) \geq 1 + N_{\rm rec}(\mathbf{x},
R)
\label{eq:zeta_barrier}
\ee is satisfied for any value of $R$. In the above expression, the
quantity $f_{\rm coll}(\mathbf{x}, R)$ is the collapsed mass fraction
and $N_{\rm rec}(\mathbf{x}, R)$ is the average number of
recombinations within the spherical volume. The parameter $\zeta$
represents the number of photons in the IGM per hydrogen nucleus in
stars and is given by \be \zeta = f_* f_{\rm esc} N_{\gamma}, \ee
where $f_*$ is the fraction of baryonic mass in stars, $f_{\rm esc}$
is the fraction of ionizing photons that escape into the IGM and
$N_{\gamma}$ is the number of ionizing photons produced per hydrogen
atom in stars. In case the condition Eq.~(\ref{eq:zeta_barrier}) is not
satisfied for any $R$, the location $\mathbf{x}$ is assigned a
ionization fraction $\zeta f_{\rm coll}(\mathbf{x}, R_{\rm cell})/[1 +
  N_{\rm rec}(\mathbf{x}, R_{\rm cell})]$, where $R_{\rm cell}$ is the
size of the grid cells.

The quantity $N_{\rm rec}(\mathbf{x}, R)$ will depend on the
small scale clumpiness in the density field and on the reionization
history in the spherical region around $\mathbf{x}$. We have already
discussed that simulating the reionization history in a cosmologically
representative volume along with resolving small scale features in the
density field is a daunting task. In this work, we circumvent this
difficulty by replacing $N_{\rm rec}(\mathbf{x}, R)$ with its globally
averaged value $\bar{N}_{\rm rec}$. This approximation does not
account for the fact that the number of recombinations will be larger
when the spherical volume under consideration is centred around a high
density region. However, the effect of this will be moderate towards
the end stages of reionization where most of the ionized regions are
so large that the average properties should be similar to the global
averaged values.

With the above approximation and the definition \be \zeta_{\rm eff}
\equiv \f{\zeta}{1 + \bar{N}_{\rm rec}}, \ee the condition for a point
to be ionized is simply given by equation (\ref{eq:zeta_eff_barrier}).
Note that we only need to specify the value of the one free parameter
$\zeta_{\rm eff}$ at each redshift for generating the ionization field
which is a combination of the properties of the halo (through the
parameter $\zeta$) and on the density structure (in particular, the
clumping factor) and ionization history of the IGM (through
$\bar{N}_{\rm rec}$).

\subsection {Ionization field in the hybrid box} 

As mentioned in Section \ref{sec:calibration}, the ionization field in
the hybrid box can be obtained by appropriately replicating the
ionization field of the small box, and then superimposing the large
scale neutral regions. Note that not all points in the IGM will have
the background photoionization rate $\Gamma_{\rm HI}$. Firstly, when $Q_M < 1$,
the large scale ionization field would not have percolated into certain regions, and hence we would have  $\Gamma_{\rm HI} = 0$ in those neutral patches.
In addition, the
self-shielding condition will make the high density regions have a
smaller photoionization rate even within ionized regions. Further, there will be spatial fluctuations in the background
photoionization rate  depending on the
distribution of sources, especially before and immediately after the
overlap of ionized regions.  In this work we ignore such
spatial fluctuations and assume the rate to be uniform within ionized
regions. The only large scale fluctuations in $\Gamma_{\rm HI}$ thus arise
from the distribution of ionized regions.

\begin{figure*}
\centering
\includegraphics[angle=0,width=0.95\textwidth]{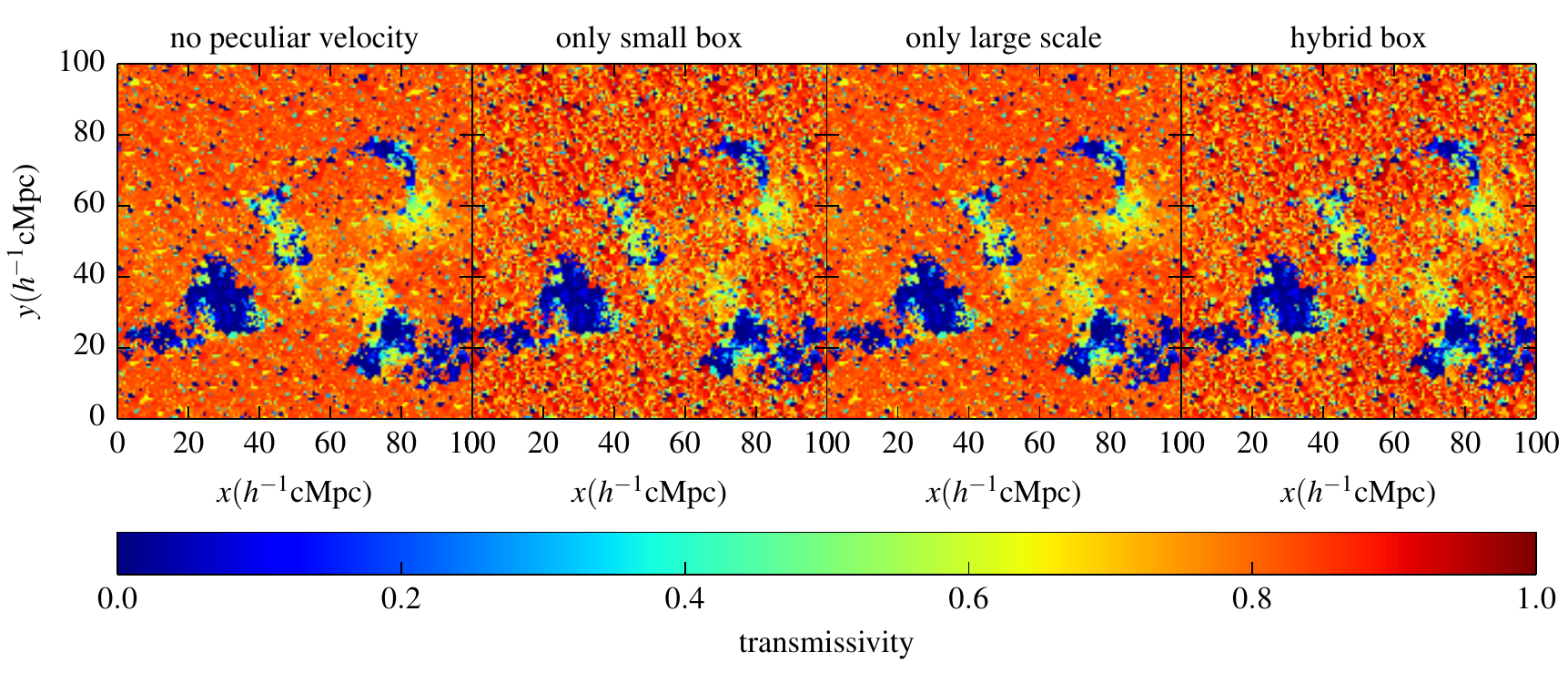}
\caption{The effects of peculiar velocity on the \lya emission line
  transmissivity. The \textit{left panel} shows the transmissivity if
  peculiar velocities are ignored. In the \textit{second} panel, we
  have accounted only for the peculiar velocities in the small
  box. The \textit{third} panel displays results for which only the
  peculiar velocities in the large box have been used. Note, that here
  only those Fourier modes in the large box that are not sampled in
  the small box were included. The \textit{right} panels shows the
  transmissivity in the hybrid box using the velocity field obtained
  by combining the velocities in the large and small boxes in Fourier
  space.}
\label{fig:plot_Tmaps_halo_velocities}
\end{figure*}

\subsection {Emissivity and other quantities of interest} 

Once the value of $\Gamma_{\rm HI}$ is known at every point in the
hybrid box, it is straightforward to compute quantities like the
emissivity $\dot{n}_{\rm ion}$, the mean free path $\lambda_{\rm mfp}$
of ionizing photons and the clumping factor ${\cal C}$.

The clumping factor ${\cal C} \equiv \overline{n_{\rm
    HII}^2}/\bar{n}_H^2$ is the most straightforward to obtain once
the ionization field in the hybrid box is set up. The mean free path
is obtained by shooting sightlines parallel to the box boundaries. We
evaluate the optical depth $\tau_{\rm HI}(x) = \sigma_{\rm HI}\int^x
\der x'~n_{\rm HI}(x')$ at the Lyman-limit $\nu = \nu_{\rm HI}$ as a
function of comoving distance $x$ along each of these sightlines. The
(comoving) mean free path is then given by \be \lambda_{\rm mfp} =
-\f{x}{\ln\avg{{\rm e}^{-\tau_{\rm HI}(x)}}}, \ee where $\avg{\cdots}$
denotes averaging over sightlines. One needs to choose the length $x$
carefully, as too large a value would imply the optically thick limit
$\avg{{\rm e}^{-\tau_{\rm HI}(x)}} \to 0$ thus the expression could
lead to inaccurate values of $\lambda_{\rm mfp}$. It is thus optimal
to choose $x$ such that we can work in the optically thin limit
$\avg{{\rm e}^{-\tau_{\rm HI}(x)}} \lesssim 1$, while still sampling a
sufficiently large path length to take into account the self-shielding
of optically thick systems. Following \citet{2013ApJ...763..146E}, we
first calculate $\lambda_{\rm mfp}$ as a function of $x$ and then
choose the optimal ray length to be the largest value of $x$ for which
$x \leq \lambda_{\rm mfp}(x)/5$. Note that our calculation of
$\lambda_{\rm mfp}$ accounts for the large scale ionization field in
addition to the small scale high density systems.

The emissivity can be computed from $\Gamma_{\rm HI}$ and
$\lambda_{\rm mfp}$ as given by Eq.~(\ref{eq:Gamma_HI}). One can then
feed in these values into Eq.~(\ref{eq:dQdt}) for $Q_M$
and verify if the assumed value of $\der Q_M/\der t$ is recovered. If
not, then one has to iteratively adjust the value of $\Gamma_{\rm HI}$
until convergence is achieved.

\section{Implementing large scale velocity fields for the transmissivity maps}
\label{app:velocity_field}

We account for the differences in the peculiar velocities of emitting
galaxies and absorbers by including information from both the small
and the large box. The peculiar velocity field in the hybrid box is
constructed by using the velocities at the appropriate position in the
small box and adding the velocity corresponding to those Fourier modes
in the large box that are not sampled in the small box, i.e. for which
$|k_x| < \pi / (10 \, \mathrm{Mpc} \, h^{-1})$, $|k_y| < \pi / (10 \,
\mathrm{Mpc} \, h^{-1})$ and $|k_z| < \pi / (10 \, \mathrm{Mpc} \,
h^{-1})$.

The effects of peculiar velocity on the \lya transmissivity are
illustrated in Fig.~\ref{fig:plot_Tmaps_halo_velocities}. In contrast
to the intrinsic line redshift due to radiative transfer effects, the
peculiar velocity of emitting galaxies and absorbers have little
impact on the transmissivity; the relative velocity
difference between host halo and absorber is typically small. The
minor changes in the transmissivity are dominated by the small scale
velocity field.

\section{Correlations between ionized regions and Ly$\alpha$ emitting galaxies}
\label{sec:correlated}

As described in the main text in  our default and late reionization models we have neglected possible spatial correlations between ionized regions and \lya emitters while evaluating the effect of  an  increased neutral mass fraction on the  visibility of observed \lya emitters. How strongly correlated the ionized regions and \lya emitters are is highly uncertain. To gauge the possible  effect of such a correlation  we have also considered an extreme model 
(the very late reionization model) where the \lya emitters are placed close to the location of massive haloes in  our large simulation box. For our simple collapse fraction model for identifying the ionized regions this results in a very strong correlation. In this model we assume that the Ly$\alpha$ emitters are hosted by
haloes of mass $> 10^{10}$ M$_{\odot}$ \citep{2010ApJ...723..869O} in the large box. Since there is no  correlation between the density fields in the
large and small boxes  in our hybrid simulation, sightlines towards  haloes in the large box would, however,  not account for  the expected larger than average number of self-shielded regions in the vicinity of the \lya emitters. Hence, for a given emitter identified in the large box, we take the sightline towards  the halo in the small box which is nearest to the massive halo in the large box under consideration. We only consider the 250 most massive haloes in the small box for this. Note that this approach still does not account for the expected similar correlation of self-shielded and ionized regions, and hence will underestimate the effect of optically thick self-shielded regions on the \lya emitter transmissivity. Reality should  lie somewhere in between our  uncorrelated and correlated models.

To match the observed evolution of \lya emitters in this model, we have found a suitable $Q_M(z)$ by trial and error
 such that the reionization is completed at $z = 6$ (as opposed to 6.7 in HM2012) and evolves rapidly at $z > 6$.  
 At high redshift  some of the properties of the very late reionization model shown as light gray points and  curves  in Figure~\ref{fig:plot_HM12} exhibit significant differences  compared to the default and late model without spatial correlations. The photoionization rate, in particular, shows a sharp rise at $z \sim 8$ due to the 
rather  low values of $Q_V$. The ionizing photons produced are  confined to smaller volumes and thus give rise to higher $\Gamma_{\rm HI}$. At the same time the emissivity in this model is lower  at high redshift to allow for a later start of  reionization, followed by a sharp increase at $z \sim 6.5$. The electron scattering optical depth is reduced to $\tau_{\rm el} = 0.058$. The resulting REW distribution, as shown in the right hand panel of Figure~\ref{fig:plot_lae_ewdist_vwind_evol}, is again  consistent with the data.

We thus see that irrespective of whether  the correlation between \lya emitters and ionized regions are modelled, it is possible to find reionization histories which are consistent with the CMB polarization data and at the same time able to explain the REW distribution of the observed \lya emitters. Note, however, that the evolution of the ionized mass fraction is  quite different in the different models, particularly towards the end of reionization.

\end{document}